\documentclass[journal=jacsat,manuscript=article]{achemso}
\usepackage{graphicx}   % Include figure files
\usepackage{natbib}
\usepackage{subfigure}
\usepackage{amsmath}
\usepackage{amssymb}
\usepackage{color}
\usepackage[version=3]{mhchem}
\usepackage{pifont}% http://ctan.org/pkg/pifont
\usepackage{array}
\usepackage{booktabs}
\usepackage{enumerate}
\usepackage{arydshln}	 % library for dashed lines in table
\usepackage{tikz}		% flowchart tools
\usetikzlibrary{shapes,arrows}
\definecolor{cellc}{RGB}{2,159,228}
\definecolor{linc}{RGB}{34,64,116}

\title{Machine learning force fields: Construction, validation, and outlook}

\author{V. Botu}
\email{botuvenkatesh@gmail.com} 
\affiliation{Department of Chemical \& Biomolecular Engineering, University of Connecticut, Storrs, CT}

\author{R. Batra}
\affiliation{Department of Materials Science \& Engineering, University of Connecticut, Storrs, CT} 

\author{J. Chapman}
\affiliation{Department of Materials Science, University of Connecticut, Storrs, CT} 

\author{R. Ramprasad}
\affiliation{Department of Materials Science \& Engineering, University of Connecticut, Storrs, CT}

\begin{document}
\begin{abstract}
Force fields developed with machine learning methods in tandem with quantum mechanics are beginning to find merit, given their (i) low cost, (ii) accuracy, and (iii) versatility. Recently, we proposed one such approach, wherein, the vectorial force on an atom is computed directly from its environment. Here, we discuss the multi-step workflow required for their construction, which begins with generating diverse reference atomic environments and force data, choosing a numerical representation for the atomic environments, down selecting a representative training set, and lastly the learning method itself, for the case of Al. The constructed force field is then validated by simulating complex materials phenomena such as surface melting and stress-strain behavior - that truly go beyond the realm of \textit{ab initio} methods both in length and time scales. To make such force fields truly versatile an attempt to estimate the uncertainty in force predictions is put forth, allowing one to identify areas of poor performance and paving the way for their continual improvement.

%Further, methods to judge force prediction accuracy are proposed allowing for an adaptive refinement of the force field. 

\end{abstract}

\maketitle

\newpage
\section{Introduction}

Materials modeling approaches largely fall in two broad categories: one based on quantum mechanical methods (e.g., density functional theory), and the other based on semi-empirical analytical interatomic potentials or force fields (e.g., Stillinger-Weber potentials, embedded atom method, etc.) \cite{Tadmor_1,Hautier_1,Burke_1,Neugebauer_1,Hill_1,Torrens_1,Elliott_1}. Choosing between the two approaches depends on which side of the cost-accuracy trade-off ones wishes to be at. Quantum mechanical methods (also referred to as \textit{ab initio} or first principles methods) are versatile, and offer the capability to accurately model a range of chemistries and chemical environments. But such methods remain computationally very demanding. Practical and routine applications of these methods at the present time are limited to studies of phenomena whose typical length and time scales are of the order of nanometers and picoseconds, respectively. Semi-empirical methods capture the essence of these interatomic interactions in a simple manner via parameterized analytical functional forms, and thus offer inexpensive solutions to the materials simulation problem. However, their applicability is severely restricted to the domain of chemistries and chemical environments intended, or considered during parameterization \cite{Bianchini_1}. It is unclear whether the underlying framework allows for a systematic and continuous improvement in the predictive capability of newer environments.

The present contribution pertains to a data-driven approach by which flexible and adaptive force fields may be developed, potentially addressing the challenges posed. By using carefully created benchmark data (say, from quantum mechanics based materials simulations) as the starting point, non-linear associations between atomic configurations and potential energies (or forces, more pertinent to the present contribution) may be learned by induction \cite{Witten_1,Hastie_1,Hoffman_1}. This data-driven paradigm, popularly referred to as machine learning, has been shown by many groups to lead to viable pathways for the creation of interatomic potentials that; (1) surpass conventional interatomic potentials both in accuracy and versatility, (2) surpass quantum mechanical methods in cost (by orders of magnitude), and (3) rival quantum mechanics in accuracy \cite{Behler_1,Bartok_2,Lorenz_1}, at least within the configurational and chemical domains encompassed by the benchmark dataset used in the training of the potential. 

A new recent development within the topic of machine learning based interatomic potentials is the realization that the vectorial force experienced by a particular atom may be learned and predicted directly given just a configuration of atoms \cite{Botu_1,Botu_2,Li_1}. This capability is particularly appealing as the atomic force is a local quantity purely determined by its local environment, in contrast to the total potential energy which is a global property of the system as a whole. A large body of materials simulations, such as geometry optimization and molecular dynamics simulations, require the atomic force as the sole necessary input ingredient \cite{Tadmor_1}. Note that partitioning the total potential energy into individual atomic contributions, conventionally adopted in semi-empirical interatomic potentials, is a matter of convenience of construction, rather than being a fundamental requirement.

\begin{figure}
	\centering
	\includegraphics[scale=0.69]{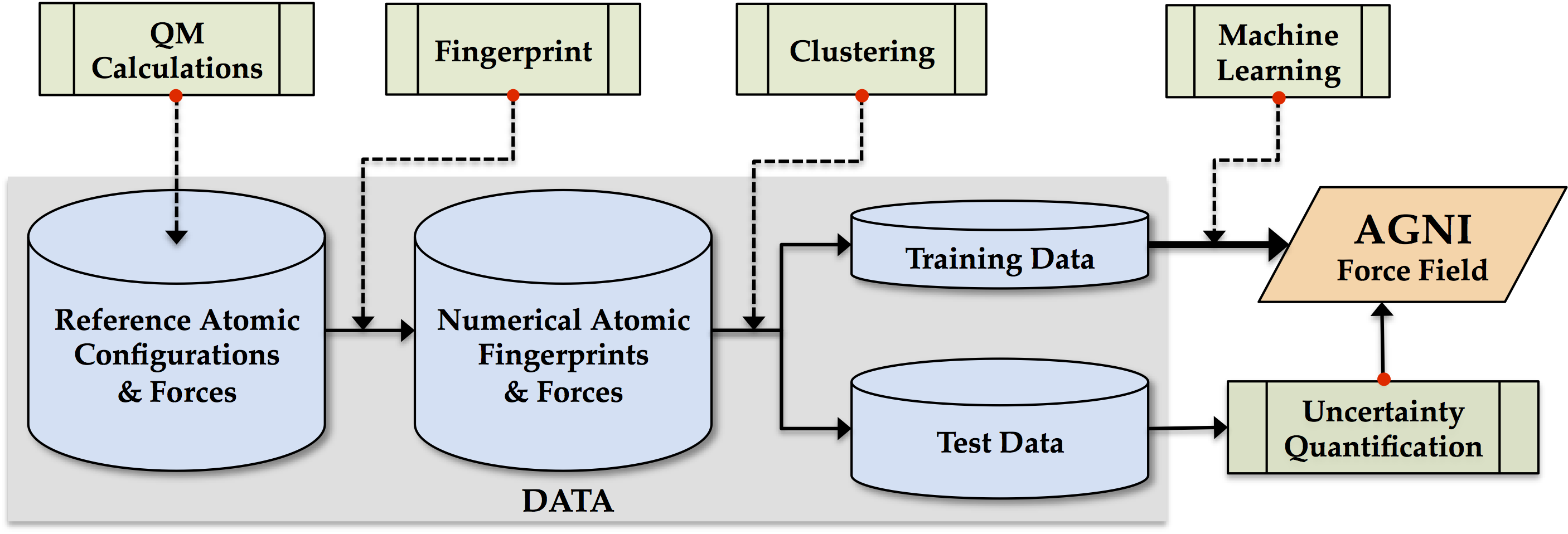}
	\caption{Flowchart illustrating the workflow in constructing AGNI force fields - generating reference atomic configurations and forces with quantum mechanical methods, fingerprinting atomic environments, rational selection of training and test datasets, mapping atomic fingerprints to forces using machine learning methods, and quantifying uncertainty in predictions made.} 
	\label{Fig:idea}
\end{figure}

This article deals specifically with using machine learning methods to create an atomic force prediction capability, i.e., a force field. As recently pointed out, this force field is \emph{\underline{A}daptive} (i.e., new configurational environments can be systematically added to improve the versatility of the force field, as required), \emph{\underline{G}eneralizable} (i.e., the scheme can be extended to any collection of elements for which reliable reference calculations can be performed), and is \emph{\underline{N}eighborhood \underline{I}nformed} (i.e., a numerical fingerprint that represents the atomic environment around the reference atom is mapped to the atomic force with chemical accuracy) \cite{Botu_1,Botu_2}. The force field is henceforth dubbed AGNI. 

The workflow in constructing AGNI force fields includes five key steps. These include: (1) creation of a reference dataset derived from a plethora of diverse atomic environments of interest and the corresponding atomic forces computed using a chosen quantum mechanical method, (2) fingerprinting every atomic environment in a manner that will allow the fingerprint to be mapped to atomic force components, (3) choosing a subset of the reference dataset (the ``training'' set) using clustering techniques to optimize the learning process while insuring that the training set represents the diversity encompassed by the original reference dataset, (4) learning from the training set, thus leading to a non-linear mapping between the training set fingerprints and the forces, followed by testing the learned model on the remainder of the dataset using best-statistical practices, and (5) finally, estimation of the expected levels of uncertainty of each force prediction, so that one may determine when the force field is being used outside its domain of applicability. The entire workflow involved in the construction of AGNI force fields is portrayed schematically in Figure \ref{Fig:idea}.

In our previous work, a preliminary version of the AGNI force field for Al was used to demonstrate its capability with respect to predicting structural, transport or vibrational properties of materials \cite{Botu_2}. Here, we further extends its scope, by including more diverse atomic environments and its ability to simulate even more complex phenomena - such as surface melting and stress-strain behavior. Furthermore, although AGNI is built to provide atomic forces, we demonstrate that accurate total potential energies can be retrieved either during the course of a molecular dynamics simulation or along a reaction coordinate, through appropriate integration of atomic forces.

Additional comments pertaining to the last step of the workflow in Figure \ref{Fig:idea} are in order. Uncertainty quantification is essential to recognize when the force field is operating outside its domain of applicability. Ideally, larger the uncertainty of the force prediction for an atom in a given environment, greater is the likelihood that the environment is ``new". By imposing a threshold and monitoring the uncertainty we may wish to augment the training set with the corresponding new atomic environment(s), and follow the workflow in Figure \ref{Fig:idea}. This helps build force fields that are truly \textit{adaptable}. Initial steps towards quantifying this uncertainty in force predictions are undertaken.

The rest of the paper is organized as follows. In the first half of the work we guide the readers through the rigors of each step in the force field construction workflow, shown in Figure \ref{Fig:idea}, to develop a general-purpose Al force field. The applicability of the force field is then validated by demonstrating its use in atomistic simulations. A discussion on measures to estimate uncertainties in force predictions made is then put forth. Lastly, we conclude with an outlook on using machine learning force fields in the field of atomistic materials modeling, and the challenges that yet remain to be addressed.

\section{Generating reference data} 

\begin{figure}
	\centering
	\includegraphics[scale=0.9]{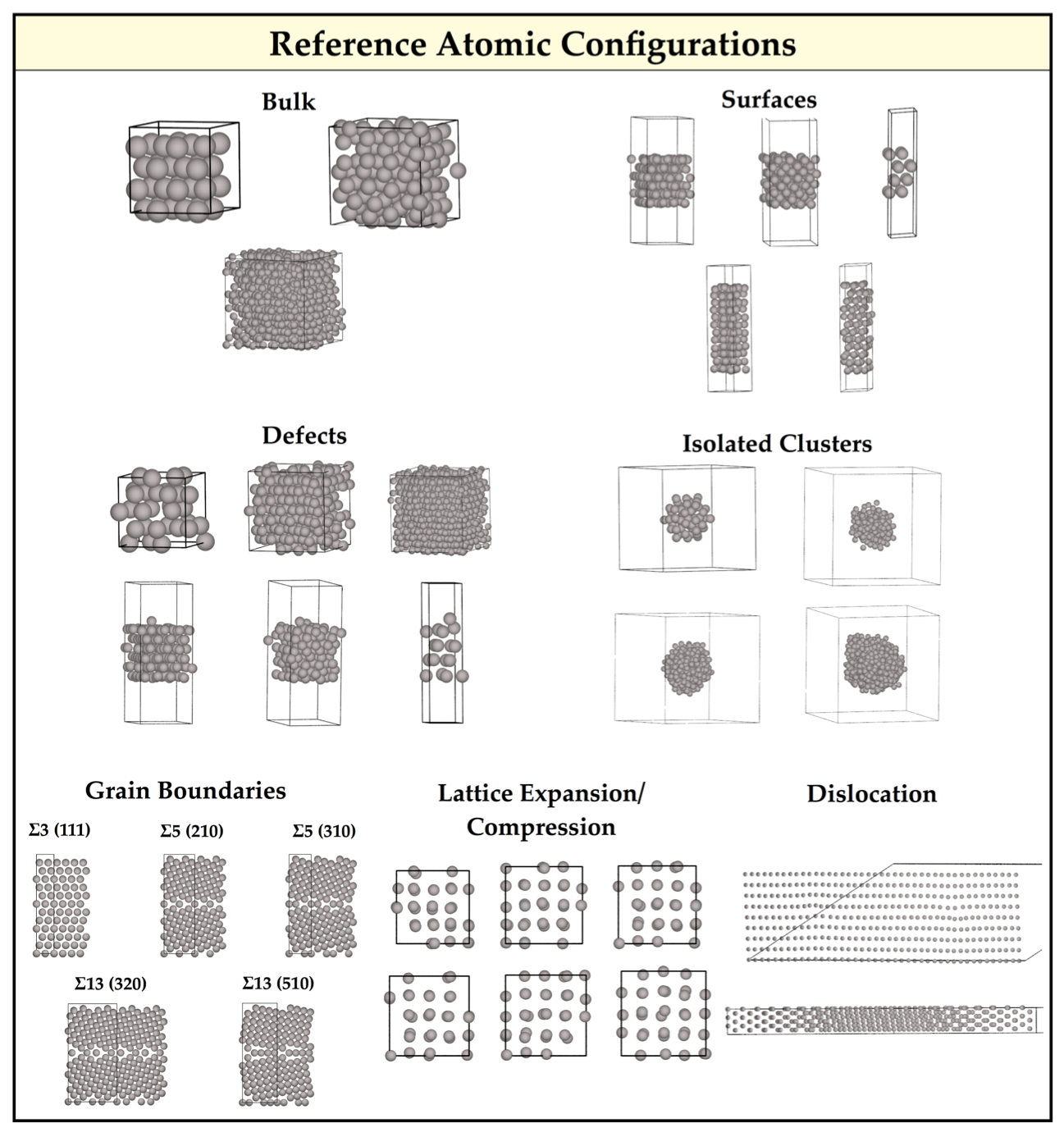}
	\caption{Reference configurations used to sample atomic environments for training and testing of AGNI force fields; (i) bulk, (ii) surfaces, (iii) defects (vacancies and adatoms), (iv) isolated clusters, (v) grain boundaries, (vi) lattice expansion and compression, and (vii) dislocation.} 
	\label{Fig:configurations}
\end{figure}

The construction of AGNI force fields begins with data. We start by building several periodical and non-periodical equilibrium configurations (c.f., Figure \ref{Fig:configurations}), such as; (i) defect free bulk, (ii) surfaces, (iii) point defects - vacancies and adatoms, (iv) isolated clusters, (v) grain boundaries, (vi) lattice expansion and compression, and (vii) edge type dislocations. These configurations are so chosen to mimic the diverse environments an atom could exist in, and forms a critical first step in constructing generalizable force fields. It is by no means a complete list and one could continuously add non-redundant configurations to it (methods to identify such redundancies are discussed later). The vectorial force components on each atom, in the equilibrium configurations amassed, are then computed by quantum mechanical based density functional theory (DFT) calculations.\cite{Kohn_1,Hohenberg_1} To correctly describe the non-equilibrium behavior of an atom, in response to a perturbation due to thermal vibrations, pressure or other sources, it is equally necessary to construct non-equilibrium atomic environments - to learn the complete array of forces experienced by an atom. A convenient and quick means to sampling such non-equilibrium environments is with \textit{ab initio} molecular dynamics (MD) simulations.\cite{Car_1} Here, starting with the equilibrium configurations in Figure \ref{Fig:configurations} constant temperature MD simulations were carried out across a range of temperatures between 200 - 800 K, resulting in a diverse set of reference atomic environments and forces (c.f., Table \ref{Table:Data_set}) - needed to learn (indirectly) the underlying potential energy surface.

\begin{center}
	\captionof{table}{Atomic environment makeup for the five datasets; A, B, C, D and E. For each dataset we generate a training and test set (except for dataset E, where only a test set is created) - the former used to construct the force field  and the later to validate it. The number of new environments added is indicated in the last column.} \label{Table:Data_set}
	\begin{tabular}{clllllc}
		\hline
		\hline
		\multicolumn{1}{c}{\textbf{Dataset}} & \multicolumn{5}{c}{\textbf{Atomic Envs. from Reference Configurations}} &
		\multicolumn{1}{c}{\textbf{Number of Envs.}}\\
		\hline
		\multicolumn{1}{c}{\textbf{A}} & \multicolumn{5}{m{10.5cm}}{Defect free bulk fcc and bcc.} &
		\multicolumn{1}{c}{20385} \\
		\hdashline
		\multicolumn{1}{c}{\textbf{B}} & \multicolumn{1}{c}{Dataset A +} & \multicolumn{4}{m{8cm}}{(100), (110), (111), (200), and (333) surfaces.} &
		\multicolumn{1}{c}{211255} \\
		\hdashline
		\multicolumn{1}{c}{\textbf{C}} & \multicolumn{1}{c}{Dataset B +} & \multicolumn{4}{m{8cm}}{Defects in bulk fcc with 1, 2 and 6 randomly distributed vacancies and adatom on (100), (110) and (111) surfaces.} &
		\multicolumn{1}{c}{1502856} \\
		\hdashline		
		\multicolumn{1}{c}{\textbf{D}} & \multicolumn{1}{c}{Dataset C +} & \multicolumn{4}{m{8cm}}{Isolated clusters of 5\AA, \ 8\AA, \ 10\AA, \ and 12\AA.} &
		\multicolumn{1}{c}{586679} \\
		\hdashline
		\multicolumn{1}{c}{\textbf{E}} & \multicolumn{5}{m{10.5cm}}{$\Sigma$3 (111), $\Sigma$5 (210),  $\Sigma$5 (310), $\Sigma$13 (320), and $\Sigma$13 (510) grain boundaries, varying lattice vectors by $\pm$7 \% of equilibrium, edge dislocation along (11$\bar{2}$) direction.} &
		\multicolumn{1}{c}{394116} \\		
		\hline
	\end{tabular}
	
\end{center} 

From within the millions of reference atomic environments collected, a subset of them are chosen as training environments to construct the force fields. The particular choice of environments plays a critical role in the generalizability of such data-driven force fields. To better understand such limits imposed by data choices we construct four datasets, labeled as A, B, C, and D, with increasing complexity and diversity of atomic environments contained (c.f., Table \ref{Table:Data_set}). For each dataset, training and test sets were created - the former used to construct the force field and the later used to validate its predictive prowess. Also a fifth dataset, E, consisting of configurations never used during force field construction (see Table 1) was created, solely to demonstrate the transferability of AGNI force fields.

All force and MD calculations were done using VASP - a plane-wave based DFT software \cite{Kresse1,Kresse2}. The PBE functional to treat the electronic exchange-correlation interaction, the projector augmented wave potentials, and plane-wave basis functions up to a kinetic energy cutoff of 520 eV were used \cite{Perdew1, Blochl1}. A 14$\times$14$\times$14 $\Gamma$-centered k-point mesh was used for the primitive Al unit cell, and scaled according to the unit cell size. A timestep of 0.5 fs was chosen for the MD simulations.

\section{Fingerprinting reference environments}
%construct the force on an atom, in 3-dimensional space, one requires any 3 non-parallel force components. Depending on the direction chosen ($\hat{u}$) the component of the net force along this direction ($F^{u}$) will vary. In order to be able to map an atom's local environment to these force components, the representation chosen should  

\begin{figure}
	\centering
	\includegraphics[scale=0.65]{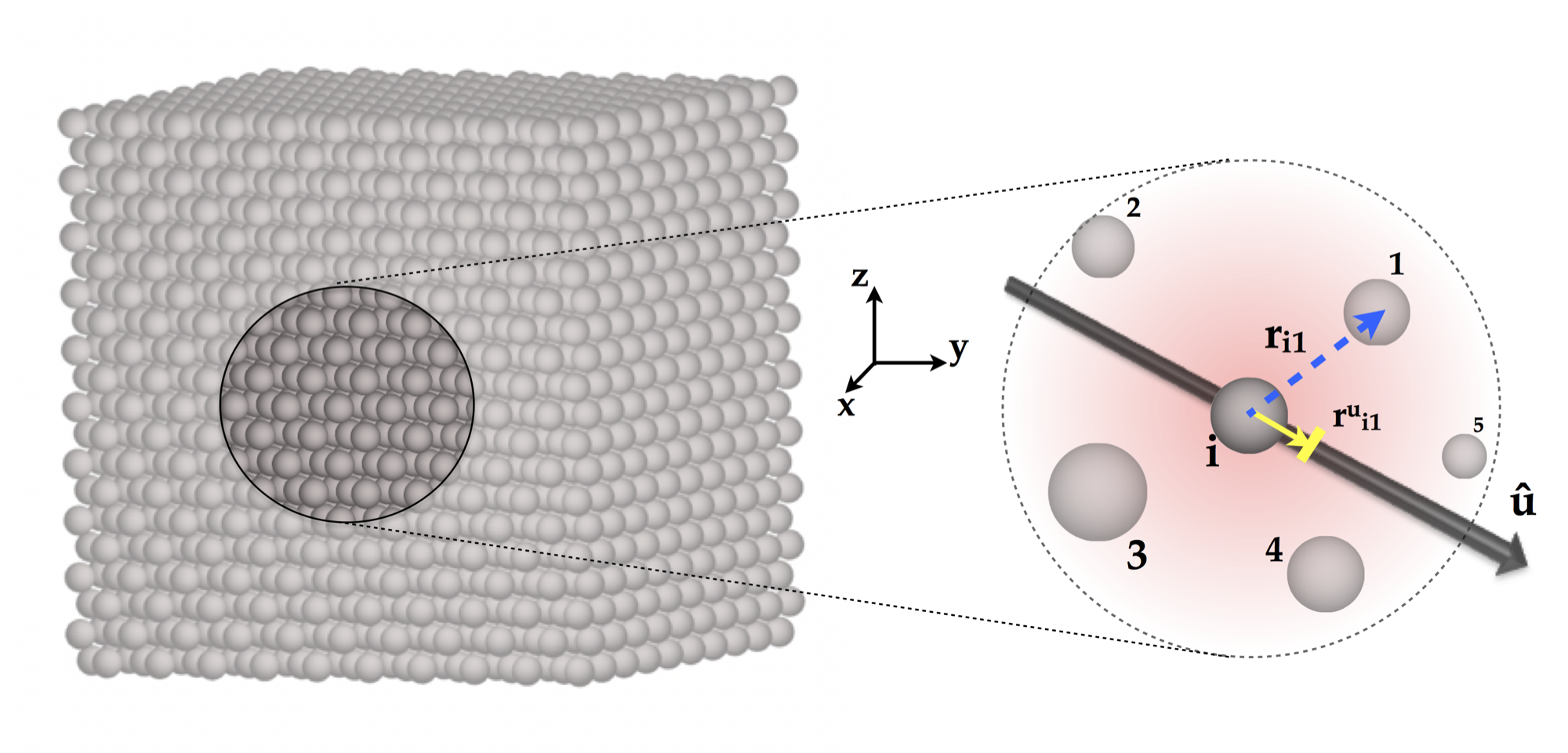}
	\caption{A schematic demonstrating the scalar projection for an atom $i$ (the reference atom) and one of its neighbor (1) along a direction $u$. To generate the final fingerprint for atom $i$, a summation over the atoms within the cutoff sphere, as indicated by the dashed line, are considered.} 
	\label{Fig:fp_projection}
\end{figure}

Choosing a representation for an atom and it's environment is the most critical step in the entire workflow. In order to learn the vectorial force components, $F^{u}$, where $u$ refers to any arbitrary direction, necessitates a numerical representation that conforms with this directional dependence. Further, it should also remain invariant to the basic atomic transformation operations, such as translation, rotation or permutation. One such representation (commonly referred to as as fingerprint) with the necessary prerequisites is given below, 

\begin{equation}\label{eq:atom}
	V_i^u(\eta) = \sum\limits_{j \neq i} \frac{r_{ij}^u}{r_{ij}} \cdot e^{-{\left(\frac{r_{ij}}{\eta}\right)}^2} \cdot  f_d{(r_{ij})}.
\end{equation}
Here, $r_{ij}$ is the distance between atoms $i$ and $j$ ($||\mathbf{r}_j - \mathbf{r}_i||$), while $r_{ij}^u$ is a scalar projection of this distance along a direction $u$ (c.f., Figure \ref{Fig:fp_projection}). $\eta$ is the Gaussian function width. $f_d(r_{ij}) = 0.5\left[\cos\left(\frac{\pi r_{ij}}{R_c}\right)+1\right]$ is a damping function for atoms within the cutoff distance ($R_c$), and is zero elsewhere. The summation in Eq. \ref{eq:atom} runs over all neighboring atoms within an arbitrarily large $R_c$ (8 \AA, in the present work). 

To better understand the fingerprint described in Eq. \ref{eq:atom}, one can deconvolute it into three sub-components (as separated by ``$\cdot$"). The exponential term ($e^{-{\left(\frac{r_{ij}}{\eta}\right)}^2}$) imposes a coordination shell around an atom $i$ with $\eta$ describes the extent of the shell. By using multiple such $\eta$ values both nearby and distant coordination information are contained within the fingerprint. In this work, $\eta$'s were sampled on a logarithmic grid between [0.8\AA, 16\AA], ensuring a sufficient description of the neighbor interactions. One could also use the peak positions of a radial distribution function as a starting point in choosing the $\eta$ values. The normalized scalar projection term ($\frac{r_{ij}^u}{r_{ij}}$) adds directionality to the fingerprint by selectively resolving the coordination information along the desired direction ($u$), and is necessary to map the individual force components. Lastly, the damping function ($f_d({r_{ij}})$) diminishes the influence of far away atoms smoothly. The combination of these three features makes this particular choice of representation suitable for mapping atomic force components. Similar coordination based fingerprints have been developed in the past \cite{Behler_1,Bartok_1}. However, these were tailored for the purpose of mapping the total potential energy (a scalar quantity) for a given configuration of atoms, unlike the vectorial force components as done here.

Further, the atomic fingerprint representation chosen conforms with the required invariance operations, such as permutation, translation, and rotation of atoms. For instance consider a reference atom, $i$, and its neighboring atoms within a cutoff sphere (as shown in Figure \ref{Fig:fp_projection}). Information pertaining to atoms neighboring atom $i$ is passed into the summand of Eq. \ref{eq:atom} as pair-wise distances, thereby, permutation or translation of atoms does not alter $V_i^u$. In the case of rotations, both the fingerprint and vectorial force components change in a manner governed by the rotation matrix. For example, by rotating atoms along the z-axis one can redefine the forces (shown here for the force but equally applicable to the fingerprint) along the Cartesian directions as,

\[
\begin{bmatrix}
F^{x'}\\
F^{y'}\\
F^{z'}
\end{bmatrix}
=
\begin{bmatrix}
\cos(\theta) & -\sin(\theta) & 0 \\
\sin(\theta) & \cos(\theta) & 0 \\
0 & 0 & 1 \\
\end{bmatrix}
\begin{bmatrix}
F^x\\
F^y\\
F^z
\end{bmatrix}. 
\]

\noindent Nevertheless, the magnitude of the net force before and after rotation remains the same, as one would expect. The concurrence to this rotational behavior implies that both the forces and fingerprints transform in an identical manner upon rotation, as needed to capture the directional behavior of forces. Another important aspect of the fingerprint is that it remains unique for the diverse atomic environment situations. Numerically this implies that identical fingerprints should map to the same atomic force value. Here, we do this by using multiple $\eta$ values. In the limit that number of $\eta$ values tends to $\infty$ uniqueness can be ensured, nevertheless, for all practical purposes one can make do with a much smaller subset, as determined by running convergence tests.

Now, using Eq. \ref{eq:atom} the atomic fingerprint along the Cartesian directions for each atom within the database is computed (since the force components obtained from DFT are along the Cartesian directions). By taking advantage of the rotational behavior, further, for each atom several arbitrary directions ($u$ in Figure \ref{Fig:fp_projection}) are defined along a spherical mesh for which the atomic force and fingerprint are reconstructed. By doing so, the pre-existing \textit{ab initio} reference database can be expanded upon with no additional costly \textit{ab initio} calculations. Though such an undertaking ensures diversity and completeness in the reference atomic environments it builds in extensive redundancies. Training a force field on millions of reference atomic environments is impractical, computationally very demanding, and might lead to misbehaved models, therefore, further down-sampling from within this big pool of data is an essential step in the construction workflow.

\section{Clustering reference data}

%\subsection{Identifying redundancies}
The next step in the construction workflow is to select a representative set of atomic environments for training purposes. To do so, it is necessary to identify the redundant and non-contributing data points from within the millions sampled. An obvious place to start is by comparing amongst the individual atomic fingerprints. However, given its high-dimensionality understanding or unraveling the fingerprint directly is non-trivial. Therefore, we rely on dimensionality reduction techniques such as principal component analysis (PCA) to project $V_i^u$ onto a lower dimension space \cite{Jolliffe_1}. 

\begin{figure}
	\centering
	\includegraphics[scale=0.23]{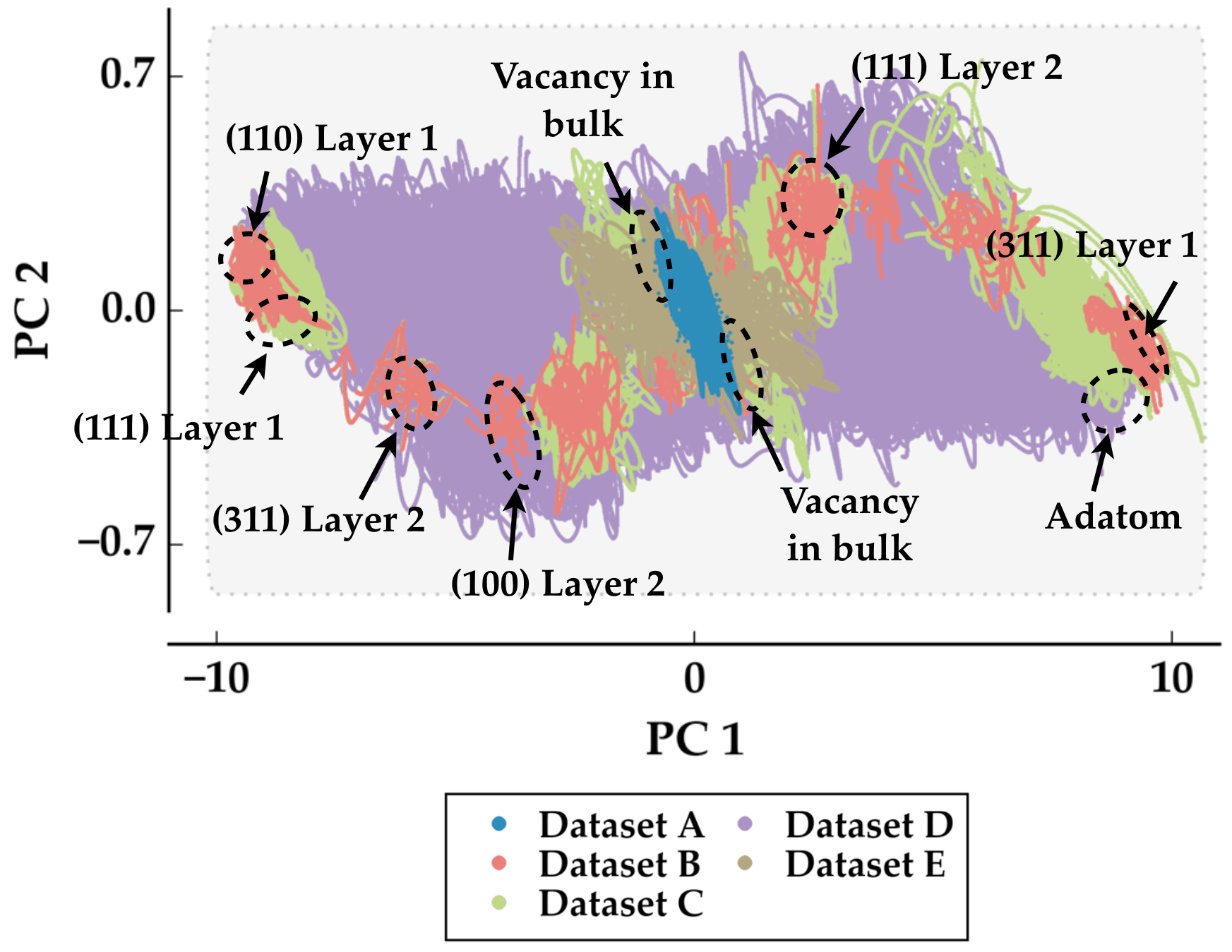}
	\caption{A projection of an 8 dimensional (i.e., 8 $\eta$ values) atomic fingerprint for all entries within dataset A, B, C, D, and E onto a 2-dimensional manifold, transformed by principal component analysis. Identical atomic environments cluster together and highlights the extent of redundancies within the reference dataset.} 
	\label{Fig:pca_projection}
\end{figure}

In PCA the original atomic fingerprint is linearly transformed into uncorrelated and orthogonal pseudo variables, also known as principal components (PCs). Often times the information content contained within the original fingerprint can be captured by a few such PCs. To demonstrate this, here, for all the reference atomic environments we compute and transform an 8-dimensional fingerprint (the rationale for which shall be discussed shortly). Two such PCs captured more than 99\% of the information content of the original fingerprint, allowing us to visualize the atomic environments on a two-dimensional manifold known as a scores plot (c.f., Figure \ref{Fig:pca_projection}). Immediately, we observe clustering of the atomic fingerprints that correspond to similar neighborhood environments. For clarity environments corresponding to a few such cases, e.g. adatoms, surfaces, vacancies, etc., are labeled. Further, by color coding atoms according to the dataset they were sampled from, i.e. A, B, C, D or E, we qualitatively observe their extent of diversity. For instance, dataset D (c.f., Figure \ref{Fig:pca_projection}) spans a diverse set of atomic environments as it populates majority of the space, suggesting that isolated cluster configurations are a good starting point to sample reference data from. Interestingly, atomic environments from dataset E lie within the domain of dataset D. This suggests that a force field trained on dataset D should accurately predict the forces for environments in dataset E, though they were never explicitly included during training.

Thus far, the dimensionality reductions methods used provided a visual means to identifying redundancies, but for an efficient force field construction an automated sampling of the non-redundant training environments is necessary. One way of doing so is to choose the data randomly. Unfortunately, this biases sampling according to the underlying probability distribution of the dataset and fails to sample sparsely populated regions. To avoid such irregularities, here we adopt a simple grid-based sampling on the PCA space. The PCA transformed data is split into uniform sub-grids, the bounds of which are determined by the minimum and maximum of the relevant PCs (which can be more than 2). Training points are then randomly sampled from within each sub-grid. By using a fine grid one can ensure uniform and diverse sampling from all regions of the PC space. Note that in the limit the grid size becomes large this approach is equivalent to the random sampling approach. Also, the test sets for validation purposes are generated in a similar manner from the remaining non-sampled data. Other dimensionality reduction algorithms, such as kernel-PCA \cite{Scholkopf_1} or multi-dimensional scaling \cite{Cox_1}, could similarly be adopted to sample for a representative dataset. Irrespective of the choice, such clustering methods are critical as the learning and prediction cost scales as $\mathcal{O}(n^3)$ and $\mathcal{O}(n)$, respectively (where $n$ is the training dataset size). 

%could use k-means clustering methods to sparsify and identify a diverse set of atomic environments, as has been done in constructing machine learning force fields to predict total energies \cite{Szlachta_1}. However, given the non-isotropic nature of our dataset k-means performs poorly. In this work, 

\section{Learning algorithm}

The next vital ingredient required in putting together a predictive framework is the learning algorithm itself. Deep learning neural networks\cite{Behler_2} and non-linear regression processes\cite{Bartok_2} have been the methods of choice for models describing atomic interactions. Their capability to handle highly non-linear relations, as is in the case of mapping an atom's environment to the force it experiences, makes them a suitable choice here as well. Here, we choose non-linear kernel ridge regression (KRR) method as the machine learning workhorse \cite{Muller_1,Hoffman_1}. KRR works on the principle of (dis)similarity, wherein, by comparing an atom's fingerprint ($V_i^u(\eta)$) with a set of reference cases, an interpolative prediction of the $u^{\textrm{th}}$ component of the force ($F_i^u$) can be made,
  
\begin{equation}\label{Eq: krr}
F_i^u = \sum_{t}^{N_t} {\alpha_t \cdot \textrm{exp}{\left[-\frac{\left({d_{i,t}^u}\right)^2}{2l^2}\right]}}.
\end{equation}

Here, $t$ labels each reference atomic environment, and $V^u_t(\eta)$ is its corresponding fingerprint. $N_t$ is the total number of reference environments considered. ${d_{i,t}^u} = ||V_i^u(\eta) - V_t^u(\eta)||$, is the Euclidean distance between the two atomic fingerprints, though other distance metrics can be used. $\alpha_t$s and $l$ are the weight coefficients and length scale parameter, respectively. The optimal values for $\alpha_t$s and $l$ are determined during the training phase, with the help of cross-validation and regularization methods. For further details concerning the learning algorithm the reader is directed to these sources \cite{Hastie_1,Hansen_1,Rupp_1}. 

Finally, in order to evaluate the performance of a developed force field, three error metrics; mean absolute error (MAE), maximum absolute error (MAX), and the standard deviation (in particular 2$\sigma$), were chosen. Relying on multiple metrics reduces any bias, unknowingly, introduced during model selection as shall be discussed shortly.

\section{Constructing the force field}

At this stage all the pieces required to construct AGNI force fields, as illustrated by the flowchart in Figure \ref{Fig:idea}, have been laid out. In the upcoming sections we discuss how one chooses the appropriate number of $\eta$ values to adequately describe an atomic environment, and the minimum number of training environments to use, as needed to construct accurate force fields. The accepted force field is then put to test by predicting forces on atoms outside the domain of training environments used, to ensure its generalizability.

\subsection{Convergence tests}
The first step to attaining an optimal force field is to ensure convergence with respect two parameters: (i) the number of $\eta$ values used for the atomic fingerprint, and (ii) the training dataset size. As mentioned earlier the number of $\eta$ values governs the resolution with which an atom's local coordination environment is described, while, the size (and choice, as shall be elaborated in the next section) of training data governs AGNI's interpolative predictive capability. In order to identify this optimal parameter set, we systematically increase the fingerprint resolution from 2 to 16 $\eta$ values and the training dataset size from 100 to 2000 atomic environments, while monitoring test set error (in this case the chosen error metric was the MAE). To remind the reader, $\eta$ values were sampled on a logarithmic grid between [0.8\AA, 16\AA], while training data environments were sampled using the PCA projection followed by a grid-based sampling. 

\begin{figure}
	\centering
	\includegraphics[scale=0.35]{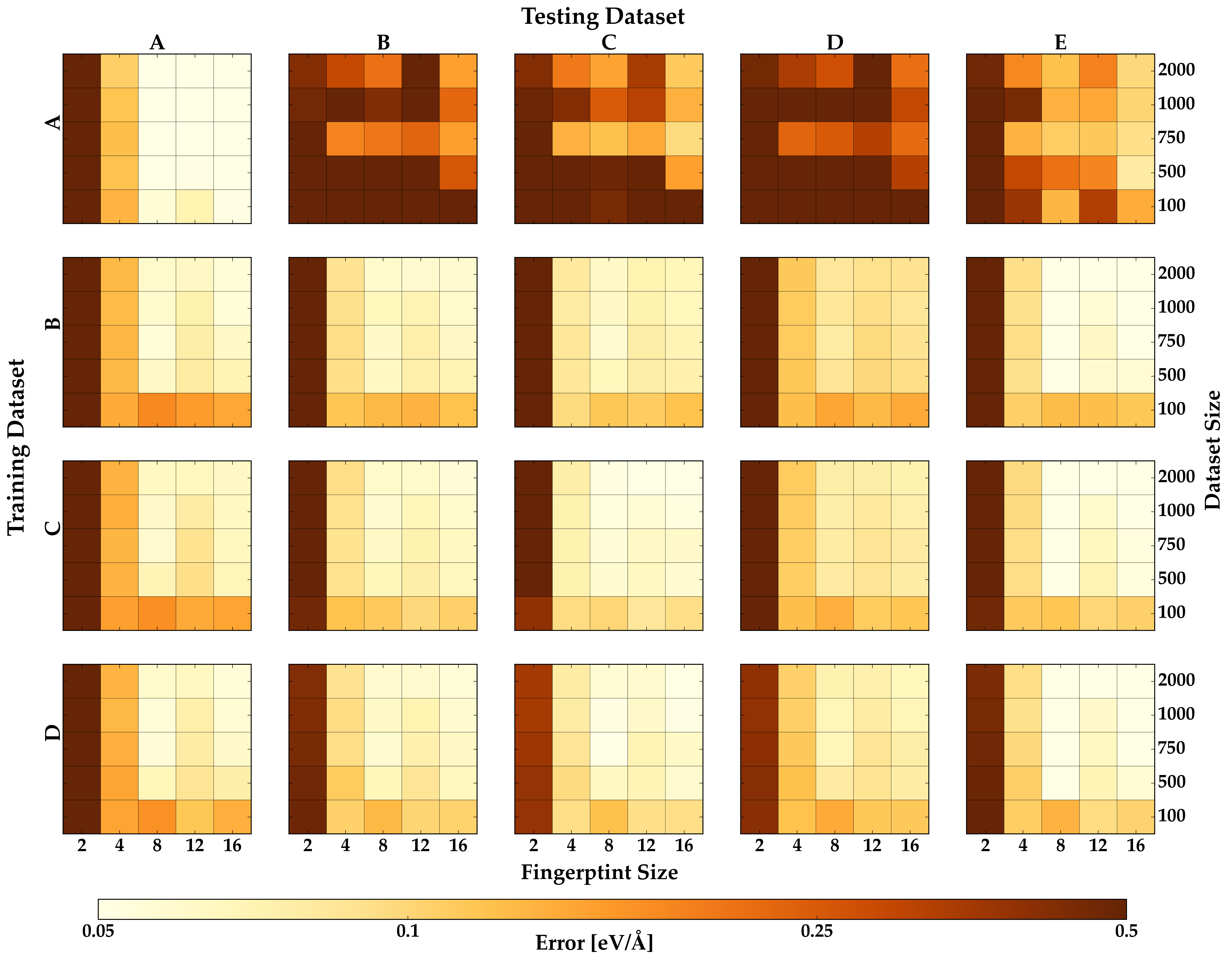}
	\caption{Heat maps illustrating model error (mean absolute error) as a function of number of $eta$ values and the training dataset size. The fingerprint was varied from 2 to 16 $\eta$ values, while the training dataset size was varied from 100 to 2000 environments. We report the MAE for models trained on datasets A, B, C, and D and tested on datasets A, B, C, D, and E. For example the top row corresponds to models trained on dataset A, while each column corresponds to a test datasets of the five cases. The errors quickly converge for a fingerprint with 8 $\eta$ values and a training size of 1000 diverse environments.} 
	\label{Fig:heatmap}
\end{figure}

For each of the four training datasets (A, B, C and D), partitioned in the data generation stage, and for all combinations of the convergence parameters an AGNI force field was constructed. Each force field is then validated on the respective test datasets (A, B, C, D and E). The force fields are denoted as $M_i^j$, where $i$ and $j$ label the training and test environments used, respectively (the superscript is omitted when referring to the training environments only). Figure \ref{Fig:heatmap} illustrates heat maps of the error for different training and test datasets, and convergence parameter combinations. Two key findings stand out: (i) by increasing the fingerprint resolution the error drops and quickly converges below $\approx$ 0.05 eV/{\AA} (expected chemical accuracy), and (ii) increasing the training dataset size reduces error only beyond a reasonable fingerprint resolution. For example, in $M_C^A$ increasing the training dataset size for a fingerprint with 2 or 4 $\eta$ values has no effect on the predictive capability. Such a manifestation implies that 8 or more $\eta$ values are required to ``uniquely'' discern amongst the atomic environments, in order for the learning algorithm to work. Nevertheless, this relation only holds for force fields used in an interpolative manner, as seen in the failure of $M_A^B$, $M_A^C$, $M_A^D$ or $M_A^E$. Here, the diversity in the training data chosen plays a more prominent role in governing performance, as shall be elaborated in the next section. Overall, we find that a fingerprint of 8 $\eta$ values and a training size of 1000 atomic environments is sufficient, beyond which the models exhibit diminishing returns, i.e. increased model training costs with no significant drop in model error, and are the parameters chosen for all subsequent discussions. The computational burden of each AGNI prediction is $\approx$ 0.1ms/atom/core, while DFT costs $\approx$ 1ks/atom/core. 

%For example, a force field generated on training environments from dataset A and evaluated on test environments from dataset C is labeled as $M_A^C$.

\subsection{Training data choice} 

\begin{figure}
	\centering
	\includegraphics[scale=0.44]{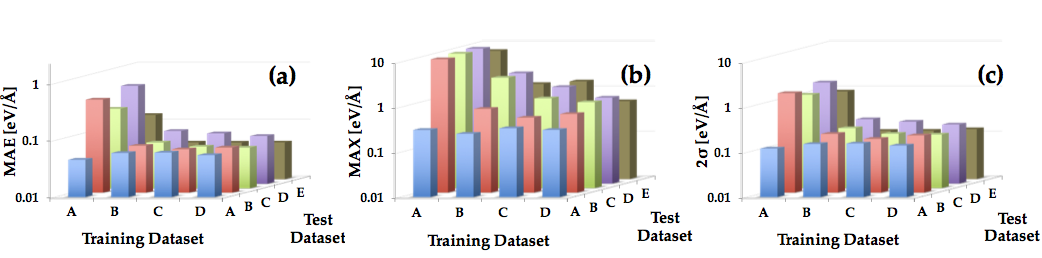}
	\caption{(a) Mean absolute error, (b) maximum absolute error, and (c) 2 * standard deviation error metric, for models trained on A, B, C and D and tested on dataset A, B, C, D and E. The errors are reported for an 8 dimensional fingerprint vector (i.e., 8 $\eta$ values) chosen to represent the environment around an atom, and a training set size of 1000 environments obtained from the PCA grid-based sampling.} 
	\label{Fig:barchart}
\end{figure}

The particular choice of atomic environments included during training is a crucial factor, as briefly alluded to earlier. Given that the learning algorithm is interpolative by nature, a force field trained say only on bulk type environments ($M_A$) cannot predict the forces corresponding to other environments types, e.g. datasets with surfaces and other features - $M_A^B$, $M_A^C$, $M_A^D$ or $M_A^E$. By increasing the diversity in training environments, $M_B$, $M_C$ and $M_D$, we make the force fields more generalizable once the optimal parameters are chosen, as given by their low test error in Figure \ref{Fig:heatmap}. Surprisingly, it appears as though predictions made with $M_B$ are equally as good as $M_C$ or $M_D$. However, this is purely a manifestation of using the MAE as the error metric. Along with the MAE, we report test set errors computed with two other metrics - MAX and 2$\sigma$, as illustrated in Figure \ref{Fig:barchart} (shown only for the optimal 8-component fingerprint and 1000 training atomic environments). For $M_B$, with MAX as the metric, the prediction error is high outside its domain of applicability (test set C, D or E), and a similar behavior is observed for $M_C$. It should be recognized that MAX reports the worst prediction made, while MAE reports a mean error skewed by test set size. By combining the two metrics with the actual variance in the errors, as measured by the 2$\sigma$ metric, we can ensure that the error is indeed under control. We observe that in $M_D$, by sampling atomic environments from a very diverse set of configurations all the error metrics are low, and the force field is highly generalizable, and is the force field used in subsequent discussions.

\subsection{Testing out of domain configurations}

\begin{figure}
	\centering
	\includegraphics[scale=0.625]{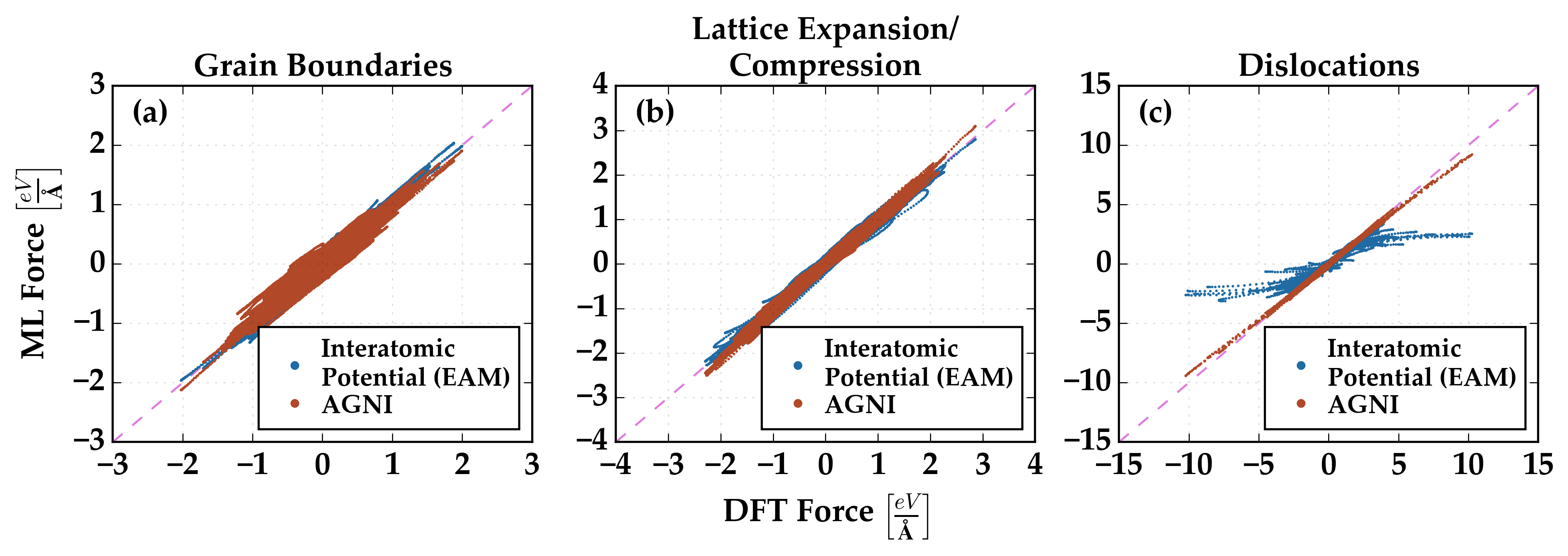}
	\caption{Parity plots comparing force predictions, for test environments in dataset E, made with the AGNI model, $M_D$, EAM interatomic potential, and DFT. The plots have been further separated according to the configuration they were sampled from, (a) grain boundaries, (b) lattice expansion/compression, and (c) edge dislocation, respectively.}
	\label{Fig:other_cases}
\end{figure}

The configurations contained in dataset E, grain boundaries, lattice expansion and compression, and dislocations, were never ``observed'' during the training phase. Being able to accurately predict the forces will further demonstrate the fidelity in using a local-neighborhood based force predictive capability. The PCA scores plot, shown in Figure \ref{Fig:pca_projection}, provided a glimpse of what one could expect. Given that the transformed atomic fingerprints for dataset E lies within the domain of environments from dataset D, one can expect the force predictions made by $M_D$ to be interpolative, and thus accurate. However, a more stringent test is to predict forces on all the atoms in dataset E and compare them to those obtained by DFT methods. As is done and shown in Figure \ref{Fig:other_cases}. For all three cases, the AGNI predicted forces are in excellent agreement with DFT. This demonstrates the intended goal of AGNI force fields, i.e. to retain quantum mechanical accuracy, be computationally inexpensive, and remain generalizable. The last feature in particular, generalizability, is often lacking with traditional semi-empirical methods. For comparison, we recompute the forces for atoms in dataset E using traditional semi-empirical potentials. Here, we particularly use an Al EAM potential\cite{Jacobsen_1}, as it accurately captures interactions in close-packed metallic type systems. As with AGNI force field, EAM methods equally predict forces accurately for grain boundaries and lattice expansion/compression but fails for dislocation type of environments.\cite{Bianchini_1}

%This once again raises an important question in the realm of force field based simulations - \emph{can one a priori judge the error in the forces predicted}? In the next section we provide one such attempt at estimating uncertainties in the force predictions made with AGNI.

%Also, it can be seen that the errors converge around 0.05 eV/\AA, similar to the numerical noise inherent within reference DFT methods. One can achieve higher accuracies in force predictions by resorting to even higher-levels of theory. 

\section{Validating the force field}

Having demonstrated a robust scheme that allows accurate atomic force predictions for a diverse set of situations, in the subsequent sections, we demonstrate the true prowess of such AGNI force fields in facilitating atomistic simulations. Our previous work provided a glimpse of such simulations, whereby, structural optimization, vibrational property estimation, and simple MD simulations of materials were undertaken.\cite{Botu_2} Here, we extend the realm of such force fields to simulate more complex atomistic phenomena, such as surface melting and stress-strain behavior. These are particularly challenging as the atoms traverse through a multitude of environments, and an accurate prediction of the forces requires undertaking the rigorous construction workflow discussed thus far. 

The simulations were carried out using the LAMMPS molecular dynamics code. \cite{Plimpton_1} The source code and force field files required to carry out the simulations are provided as supplemental files.

\subsection{Melting behavior of an Al (111) surface}

\begin{figure}
	\centering
	\includegraphics[scale=0.75]{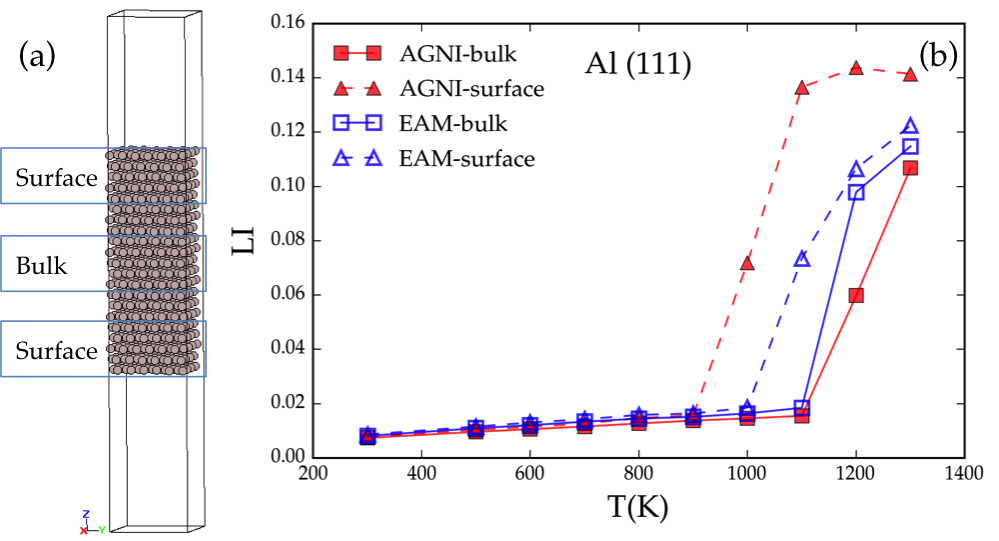}
	\caption{(a) A schematic of the (111) Al surface model and the regions classified as surface or bulk atoms. (b) The Lindemann index (LI) as function of temperature simulated with both the AGNI and EAM force fields. Melting occurs once the LI transitions from a slow linear increase to a sudden spike. With the AGNI force field the surface begins to melt $\sim$ 950 K, and propagates to the bulk by $\sim$ 1100 K.}
	\label{Fig: melting}
\end{figure}

%Here, using the constructed AGNI force field ($M_{D}$), we simulate (and then estimate) the melting temperature for an Al (111) surface.

The melting temperature of condensed matter is a property often estimated by MD simulations. Here, starting with a surface model with over 1000 atoms and dimensions of 19 {\AA} $ \times $ 17 {\AA} (c.f., Figure \ref{Fig: melting}(a)), constant temperature MD simulations were carried out for over 50 ps and across a temperature range of 300 - 1300 K, to estimate the melting temperature. In traditional MD simulations energy is used as the metric to distinguish between a solid and liquid state. Since this metric is not at our disposal we rely on the Lindemann Index (LI) order parameter instead. The LI measures the thermal perturbations of atoms - for solids this value is around $\approx$ 0.03 while for liquids it is $\approx$ 0.13. A sudden increase in the LI as a function of temperature is attributed to a solid-liquid phase transition, and can thus be utilized in MD simulations to estimate the melting temperature. \cite{Saman_1,Erik_1,Zhang_1} The LI is defined as,

\begin{equation}\label{Eq: lindemann}
LI = \frac{1}{N}\sum_{i}\frac{1}{N-1}\sum_{j \neq i}\frac{\sqrt{\langle{r_{ij}^2}\rangle - {\langle{r_{ij}}\rangle}^2}}{{\langle{r_{ij}}\rangle}}
\end{equation}

\noindent where $r_{ij}$ is the distance between atom $i$ and $j$, $N$ is the total number of atoms, and $\langle..\rangle$ indicate time averaged quantities \cite{Zhou_1}. For the range of temperatures considered the LI was computed using Eq. \ref{Eq: lindemann} and reported in Figure \ref{Fig: melting}(b) (red lines). Further, we distinguish between the surface and bulk LI values (c.f., \ref{Fig: melting}(a)) as it is well known that the melt front initiates at the surface and propagates inwards. Starting at 300 K, the LI rises due to a systematic increase in thermal vibrations up to 900 K. Between 900 - 1000 K, the sudden increase in LI signifies the onset of surface melting, which then propagates into the bulk by 1200 K. With AGNI force fields this onset of melting is observed  at $\sim$ 950 K, similar to the known experimental value at $\sim$ 933 K for Al.\cite{Stoltze_1} Even though the training environments used in building $M_{D}$ did not explicitly contain Al in a liquid state, by including high temperature MD reference data we were able to predict forces for environments in such extreme conditions. Similar LI curves computed by an EAM force field, as pure \textit{ab initio} studies of this size and timescale are intractable, yielded an overestimated melting temperature of $\sim$ 1100 K.

\subsection{Stress-strain behavior of an Al (001) surface}

\begin{figure}
	\centering
	\includegraphics[scale=0.65]{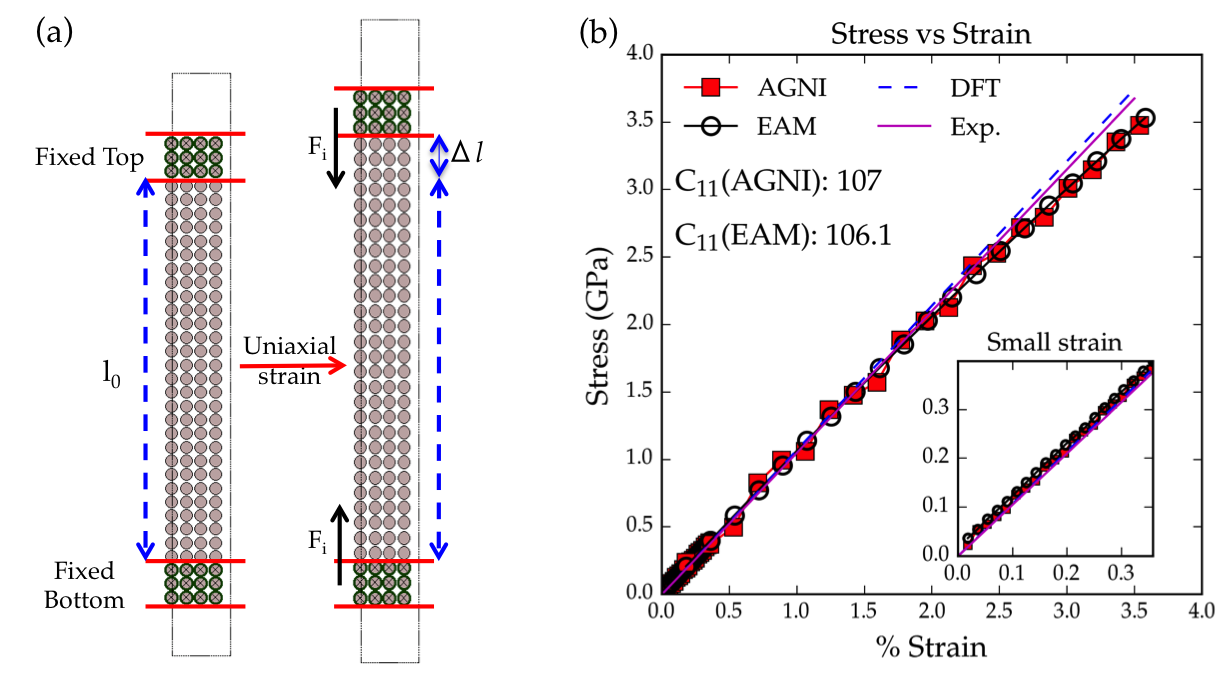}
	\caption{(a) A schematic of the surface model and procedure adopted to extract stress-strain behavior. (b) Stress vs. strain behavior of uni-axially strained Al as computed using forces predicted by the AGNI and an EAM force field. The inset shows stress at low strain ($<$ 0.35 $\%$).}
	\label{Fig: strain}
\end{figure} 

%We now demonstrate the prediction of such properties for Al using the AGNI force field ($M_{D}$).

Another important material property determined by atomistic simulations is the stress-strain behavior. Beginning with a (001) fcc Al surface, of dimensions 8 {\AA} $ \times $ 8 {\AA} $ \times $ 80 {\AA}, the surface atoms (fixed in their position) are displaced, $\Delta l$, resulting in a uniaxial strain along the surface normal [001] as illustrated in Figure \ref{Fig: strain}(a). Atoms within the strained region are then relaxed to minimize the forces acting upon them. This imposes a net force ($F_s$) on the fixed surface atoms towards the bulk (c.f., Figure \ref{Fig: strain}(a)). Using the stress tensor relation, $\sigma_{s,p} = \frac{F_s}{A_p}$, for a plane $p$ with an area $A_{p}$, across varying values of strain, $\epsilon_{s,p} = \frac{\Delta{l}}{l_0}$, one can deduce the stress-strain behavior of the material. In Figure \ref{Fig: strain}(b) we report the computed stress for varying strain deformations. The slope of this curve, for the direction considered, yields the C$_{11}$ elastic coefficient, a property that can be compared with atomistic theories. Using the AGNI force field we report a C$_{11}$ value of 107 GPa, which is in good agreement with a past \textit{ab initio} result of 105 GPa. \cite{Pham_1} Clearly, this suggests that besides forces the force field can predict their derivatives, i.e., the stresses, at quantum mechanical accuracy as well. Further, we recomputed the C$_{11}$ value with an EAM potential, resulting in a value of 106 GPa. Though, the combination of AGNI force prediction and stress relation result in quantum mechanically accurate elastic coefficients, the procedure laid out can only describe the stress along non-periodical directions, a limitation of the force based implementation.

\subsection{Energy prediction}

We now briefly touch upon the topic of energy. Energy is a unique, and important, global quantity describing the state of a configuration of atoms. It is often used to ensure stable MD simulations, estimate phase diagrams, compute minimum energy reaction pathways, etc. Given the principium of AGNI, whereby, the force on an atom is learnt based on its environment deprives the means to predicting energy directly. Nevertheless, we now discuss some alternative strategies to estimating energy, be it in dynamic or static simulations. 

\subsubsection{During a molecular dynamics simulation}

The rate of change of the total potential energy, in an MD simulation, can be expressed as a function of the individual atomic forces and velocities by invoking the chain rule,
\begin{equation}\label{Eq: energy_chain_rule}
\frac{dE}{dt} = \sum_{i,u}\frac{\partial{E}}{\partial{r_i^u}}\frac{\partial{r_i^u}}{\partial{t}} = -\sum_{i,u}F_i^u v_i^u.
\end{equation}

\noindent E is the total potential energy of the system, $r_i^u$ and $v_i^u$ are the position and velocity of atom $i$ along one of the three coordinates, $u$ $\subset{\left(x, y, z\right)}$. In the limit that $\partial{t}$ $\rightarrow$ 0, in Eq. \ref{Eq: energy_chain_rule}, an analytical expression for the rate of change in energy for infinitesimally small time difference ($\Delta$t), albeit from the initial configuration, can be expressed as,
\begin{equation}\label{Eq: energy_chain}
E_{t} = E_{t-\Delta{t}} - \Delta{t} \left(\sum_{i,u}F_i^uv_i^u\right).
\end{equation}
\noindent In MD simulations this translates to choosing a small time step to ensure accurate force integrations, and minimize numerical noise propagation, providing a pathway to indirectly monitor the energy evolution during the course of the simulation. 

\begin{figure}
	\centering
	\includegraphics[scale=0.9]{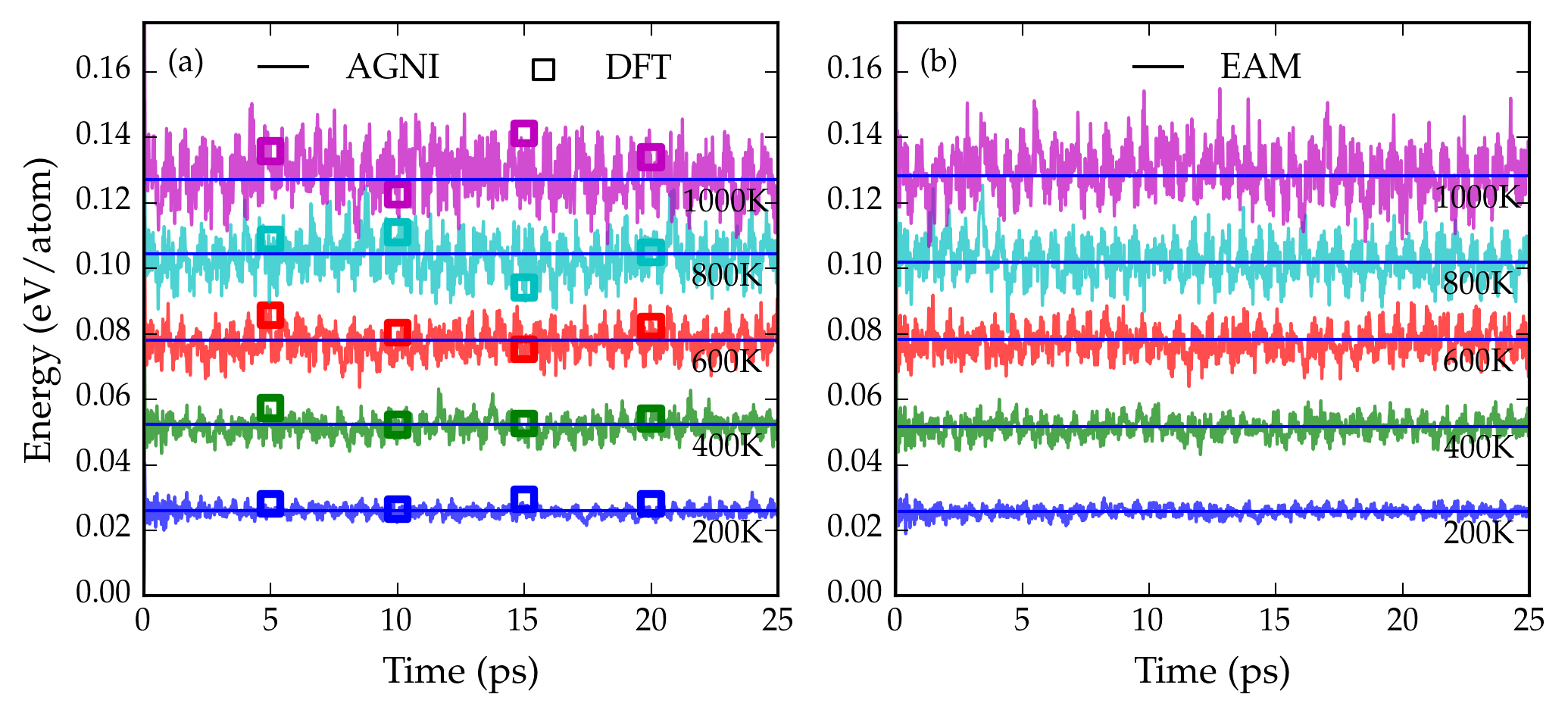}
	\caption{Evolution of energy of bulk fcc Al using the (a) AGNI with Eq. \ref{Eq: energy_chain}, and (b) EAM potential. DFT computed energies of a few configurations are also marked in (a) for validation. In all the cases, the energy is in reference to that of the perfect fcc Al. The energy is conserved with time and maintains the correct ordering with temperature.}
	\label{Fig: E_conserve}
\end{figure}

To validate this scheme, constant temperature MD simulations of bulk fcc Al ($\sim$ 250 atoms) were carried out at different temperatures using the $M_{D}$ AGNI force field. A timestep of 0.5 fs was chosen. Using the reference forces and velocities, along with Eq. \ref{Eq: energy_chain}, the rate of change in energy was computed as a function of time. The results are plotted in Figure \ref{Fig: E_conserve}(a). Clearly, the computed energies (from AGNI predicted forces) is conserved in time and maintains the correct ordering as a function of temperature. Note that the change in energy is reported with reference to the starting configuration. For comparison DFT energies (and scaled with respect to the starting configuration) for snapshots of atomic configurations along the dynamic trajectory are also reported. The error between the computed and \textit{ab initio} predicted energies is $\approx$ 4 meV/atom, close to the order of numerical noise that one can expect in such simulations. For completeness, we also report the change in energy for MD simulations run with an EAM force field (Figure \ref{Fig: E_conserve}(b)). All 3 methods display similar trends suggesting that the dynamics undertaken with an AGNI force field does indeed concur with the thermodynamic driving forces in a system.

%The potential origins of this error include; 1) inherent error in the AGNI force prediction, 2) numerical error due to the choice of $\Delta{t}$ in Eq. \ref{Eq: energy_chain}, and 3) precision error in the numerical values of the forces and the velocities. It should,however, be noted that the current AGNI framework provides a pathway to systematically improve all of these errors by utilizing the adaptive nature of AGNI, choosing a smaller timestep and increasing the numerical precision of forces and velocities. 

\subsubsection{During a static simulation}
A second, and equally simple, approach to estimating the potential energy directly from forces is by integrating them using a Taylor series approximation of the potential energy
\begin{equation}\label{Eq: Taylor}
\begin{aligned}
E = E_o + \sum_{i,u}{\left(\frac{dE}{dr^u_i}\Bigr|_{\substack{r^u_i=r_o}}(r^u_i - r_o) + \frac{1}{2!}\frac{d^2E}{{dr^u_i}^2}\Bigr|_{\substack{r^u_i=r_o}}(r^u_i - r_o)^2 + ....\right)} \\ 
\approxeq E_o - \sum_{i,u}{F_i^u \Delta{r_i^u}}, u \subset{\left(x, y, z\right)}
\end{aligned}
\end{equation}

%\begin{equation}\label{Eq: Taylor_simple}
%E \approxeq E_o - \sum_{i,u}{F_i^u \Delta{r_i^u}}
%\end{equation}

\noindent Here, $E$ is once again the total potential energy, which to a first order approximation can be derived from the atomic forces. $r^u_i$ is once again the atomic position. A discretization along the atomic positions governs the accuracy by which we can predict the energy. For these reasons Eq. \ref{Eq: Taylor} is particularly more suited for static simulations, e.g. computing reaction barriers along a reaction coordinate, when the velocities are zero. The validity of this approach was demonstrated to be consistent with the underlying potential energy surface in our previous work, whereby, the migration energy for a vacancy in bulk Al was within 3\% of the DFT predicted value. \cite{Botu_2}

Eqs. \ref{Eq: energy_chain} and \ref{Eq: Taylor} both provide a restricted means to computing the energy, whereby, it is necessary that a pathway connecting the different configurations in phase space (either in time or along a reaction coordinate) exists, in order to accurately carry out force integration. This is a limitation of using a truly force based force field, wherein, one cannot predict energies by simply choosing two arbitrary points in the phase space. Nevertheless, these findings ascertain that Eq. \ref{Eq: energy_chain} and \ref{Eq: Taylor} can indeed be used to compute change in the total potential energy as a function of time or reaction coordinate, as needed by a majority of atomistic simulations.

\section{Uncertainty quantification}

The final component to a successful predictive model is to be able to quantify the error ($\varepsilon$) in the force predictions made. If this uncertainty can be estimated, \textit{a priori}, confidence estimates for the force predictions given a new atomic environment can be provided. At the same time, it allows one to understand the force field's domain of applicability, providing a pathway for their subsequent and continual improvement.  Below is one such attempt to quantify these uncertainties.

%Further, it is a necessary and integral piece of the workflow

\begin{figure}s
	\centering
	\includegraphics[scale=0.7]{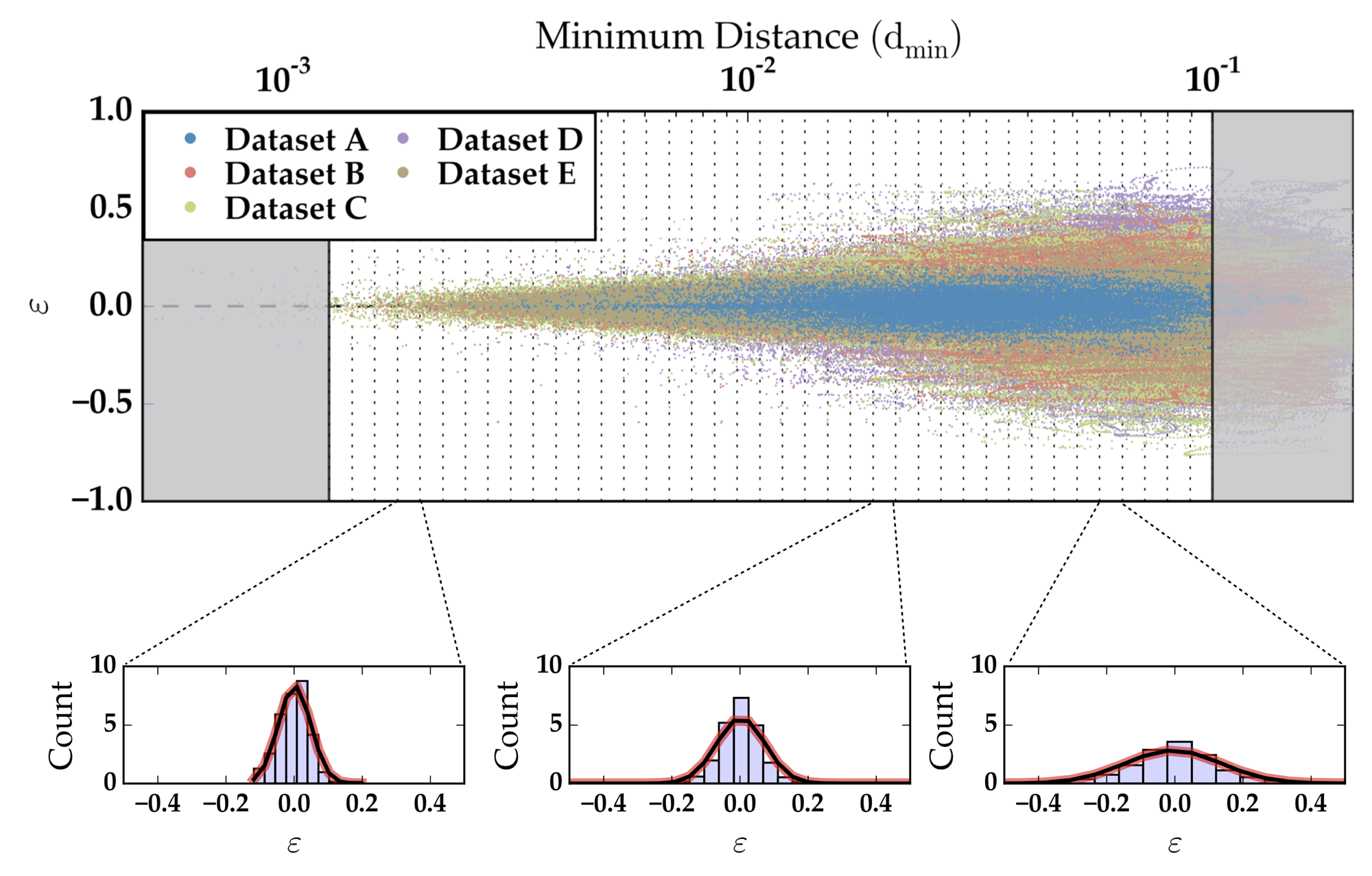}
	\caption{Top panel: a scatter plot of the minimum distance ($d_{min}$) vs. the predicted force error ($\varepsilon$). The range of $d_{min}$ is further sub-divided into small groups for statistical analysis. The gray regions were not considered for any statistical purposes, due to the lack of sufficient data (left) and high errors (right). Bottom panel: a standard normal distribution fit for each sub-group (though only shown for three such bins), used to estimate the variance in model errors.} 
	\label{Fig:uncertainity}
\end{figure}

\begin{figure}
	\centering
	\includegraphics[scale=0.85]{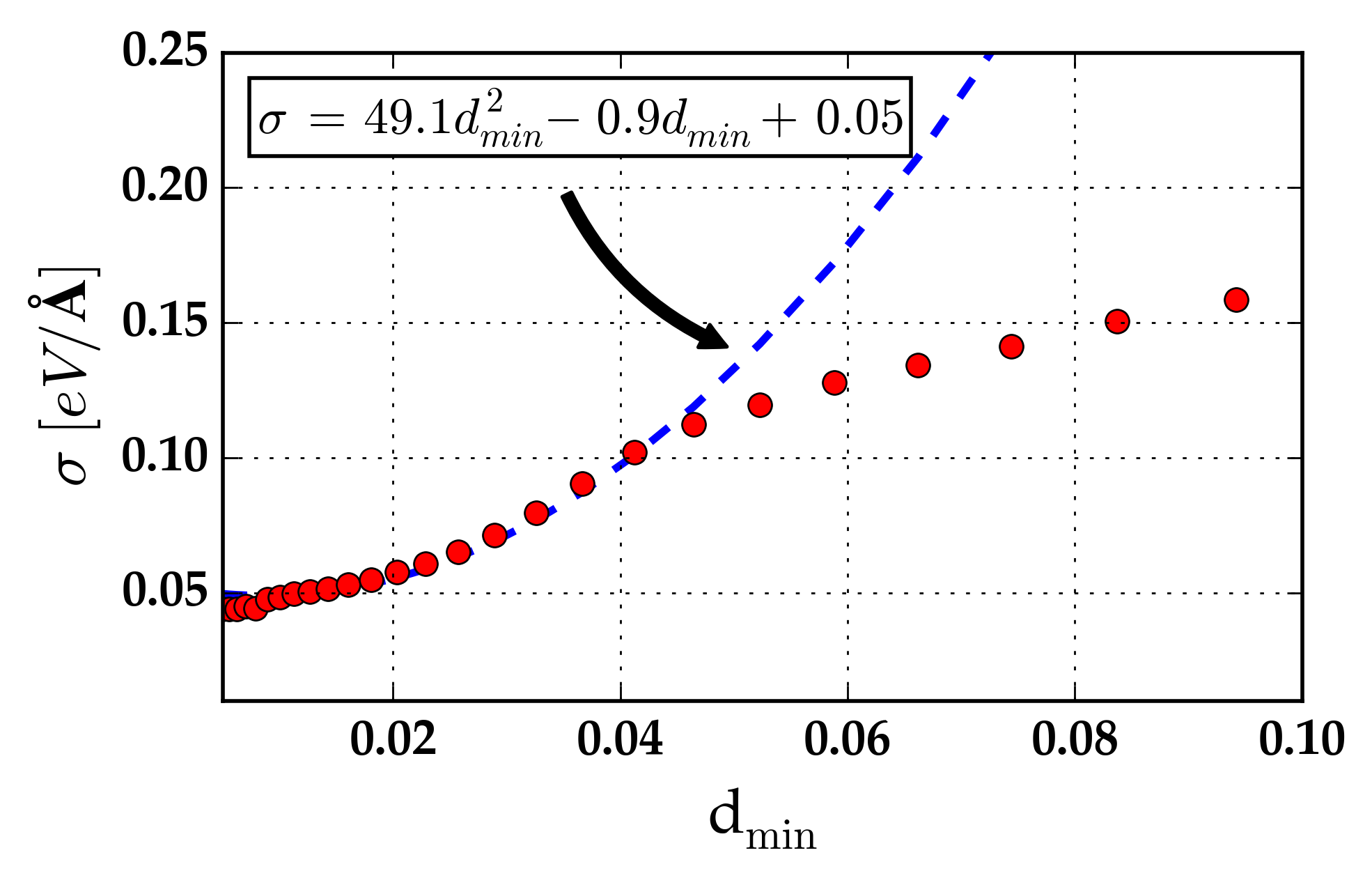}
	\caption{The uncertainty model, created for force field $M_D$, whereby $d_{min}$ is used as a descriptor to measure the expected variance in the prediction made. The markers show the actual behavior, while the blue dashed line indicates a polynomial fit to the uncertainty.} 
	\label{Fig:dmin_sigma}
\end{figure}

To compute the force on an atom, within the learning framework (c.f., Eq. \ref{Eq: krr}), begins by calculating the distance between its fingerprint and the reference training fingerprints, resulting in a total of $N_t$ distances. The final prediction is then a weighted sum of these distances, making them an important metric on predictive accuracy. Amongst the list of distances the minimum distance, $d_{min} = min \left\{d_1, d_2, . . . . ., d_{N_t}\right\}$, in particular provides a measure of ``closeness" of the new observation compared to the reference cases, and can be thought of as a descriptor in estimating $\varepsilon$. To capture this hypothesis, the $d_{min}$ and $\varepsilon$ for every observation in the test dataset was computed with the constructed AGNI force field, $M_D$. The results are summarized in the scatter plot of Figure \ref{Fig:uncertainity}. Clearly, as $d_{min}$ increases the variance in $\varepsilon$ increases - suggesting that for instances far away from reference training environments the interpolative performance fails. By binning the range of $d_{min}$s observed into uniform and smaller sub-groups, a standard normal distribution is fit to the observed $\varepsilon$. The histogram insets in Figure \ref{Fig:uncertainity} demonstrate this for three such bins. During the binning process, $d_{min}$ $<$ 10$^{-3}$ and $>$ 10$^{-1}$ were ignored due data scarcity and large predictive errors, respectively (c.f., gray-shaded regions in Figure \ref{Fig:uncertainity}). Collecting statistics across the range $d_{min}$s allowed us to fit a polynomial relation between standard deviation ($s$) and $d_{min}$, providing an analytical form to quickly estimating uncertainties. The exact functional form is $s = 49.1d_{min}^2 - 0.9d_{min} + 0.05$. This is illustrated by the red circle markers and the blue dashed line in Figure \ref{Fig:dmin_sigma}. Note that by using $s$ the confidence level in the uncertainties provided is at 68.2\%, though one can use higher confidence levels such as 2$s$ and beyond. A call that the user needs to make depending on the need and availability of computational resources.

\begin{figure}
	\centering
	\includegraphics[scale=1.0]{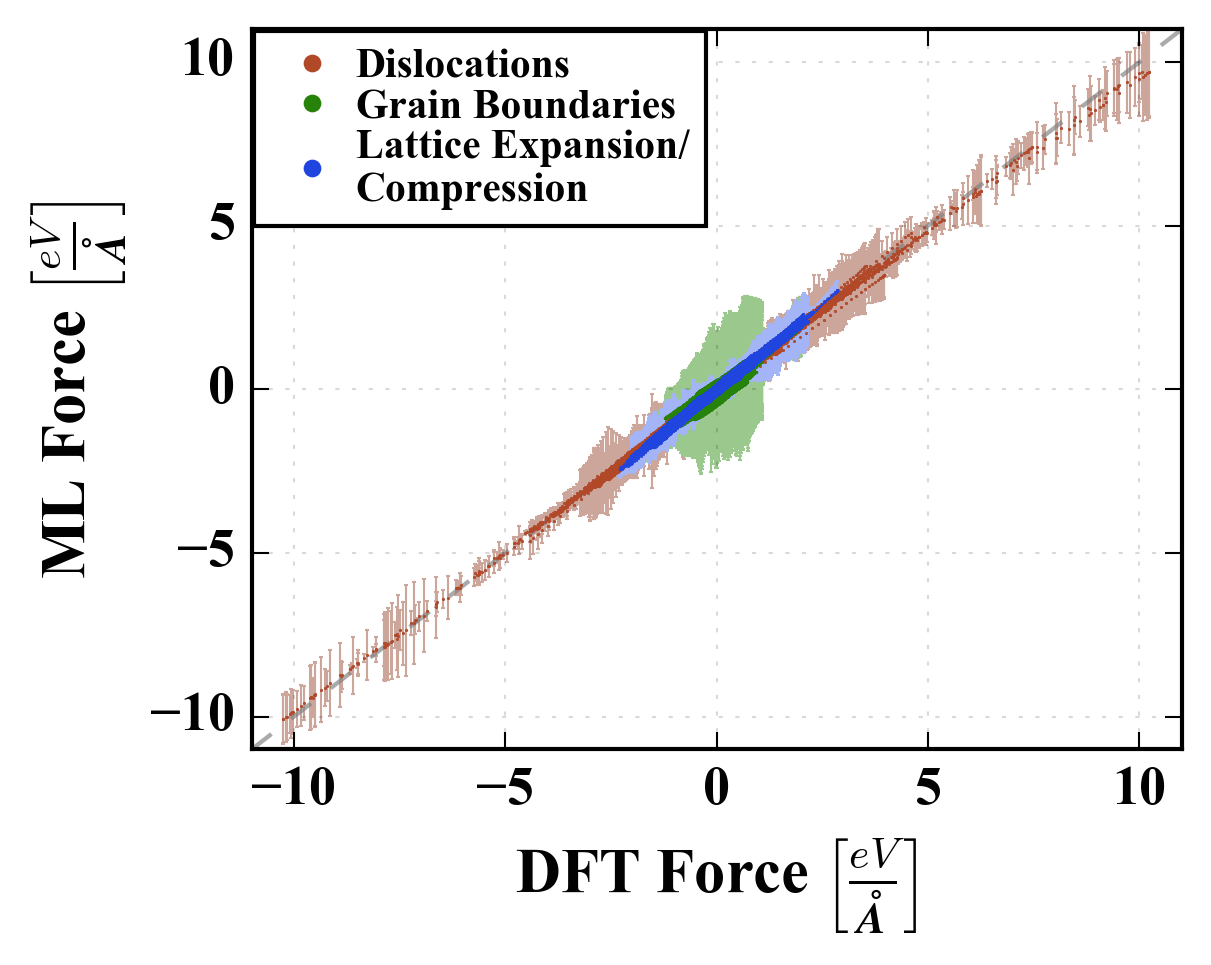}
	\caption{Parity plot of force predictions for test environments (grain boundaries, lattice expansion/compression, and dislocation) of dataset E, predicted by DFT and the ML model, $M_D$. These values are identical to those in Figure \ref{Fig:other_cases}. For each prediction the uncertainty is highlighted, as indicated by the error bars (color coded according to the configuration subclass).}
	\label{Fig:other_cases_unc}
\end{figure}

Now to demonstrate this uncertainty model, for each atom in the validation configurations (grain boundaries, lattice expansion/compression, dislocation) we estimate the uncertainty in the force predictions made. In Figure \ref{Fig:other_cases_unc} we re-plot the reference DFT and ML force predictions, along with the corresponding uncertainty in each prediction as highlighted by the error bars (color coded according to the configuration subclass marker). Immediately, those atomic environments with high uncertainties is evident and can now be flagged. These environments can be accumulated and used to retrain the force field. The preliminary steps undertaken here allows for identifying regions of poor force field performance in a systematic manner, and is integral to the continual improvement in accuracy and generalizability of AGNI force fields - making them truly \textit{adaptive}.

\section{Outlook and Summary}

A new machine learning framework to circumvent the accuracy, cost, and generalizability issues facing current atomistic models has been proposed. By directly mapping quantum mechanical derived force components to the numerical representation of an atom's local environment, accurate and computationally inexpensive force fields, herein called AGNI, were developed. In this manuscript a workflow for their systematic construction, which includes generating reference data, representing the atomic environments with a numerical fingerprint, sampling non-redundant data, learning the forces, were all demonstrated for the example of elemental Al. Further, methods to quantify uncertainties in the force predictions are proposed. This is crucial to understanding the domain of applicability of such data-driven methods, in turn paving the way for their adaptive refinement.

Nevertheless, to make such methods a mainstream tool for atomistic simulations a few challenges yet remain that need to be addressed. Firstly, to explore diverse chemistries it is necessary to come up with AGNI force fields for multi-elemental systems in an equally quick and rational manner. Though the framework discussed here was for an elemental system, the recipe is directly transferable to multi-elemental situations. Secondly, as materials science or chemical systems become ever increasingly complex, the configuration space to be explored will increase exponentially. This poses a challenge for the non-linear regression learning algorithm proposed here, and for a continued realization of machine learning force fields adopting methodologies, wherein, large quantities of data can be handled will be required. 
		
Irrespective of these challenges, the prospect of using AGNI force fields as a tool to accelerate atomistic simulations is indeed very promising. Access to such high fidelity force predictions at a fraction of the cost has already made significant in roads to studying materials and chemical phenomena. Our previous work demonstrated an expose of some atomistic simulations, such as; geometry optimization of atomic structures with several 100s of atoms, dynamical evolution of defects over long time scales (vacancies and adatoms) to determine diffusion barriers, computing vibrational properties of materials, and estimating reaction energy barriers, all using AGNI force fields \cite{Botu_2}. Here, we further extended the scope of such force fields by simulating even more complex phenomena - estimating the melting and stress-strain behavior of Al surfaces. Also, methods to reconstruct energies entirely from forces were proposed. The force field construction workflow put in place here allowed us to study these more complex materials and chemical phenomena, and such strategies are only going to become increasingly important in pushing the envelope of atomistic simulations.

\section{Acknowledgment}
This work was supported financially by the Office of Naval Research (Grant No. N00014-14-1-0098). The authors would like to acknowledge helpful discussions with K. B. Lipkowitz, G. Pilania, T. D. Huan, and A. Mannodi-Kanakkithodi. Partial computational support through a Extreme Science and Engineering Discovery Environment (XSEDE) allocation is also acknowledged.

\bibliography{references}

\providecommand{\latin}[1]{#1}
\providecommand*\mcitethebibliography{\thebibliography}
\csname @ifundefined\endcsname{endmcitethebibliography}
  {\let\endmcitethebibliography\endthebibliography}{}
\begin{mcitethebibliography}{41}
\providecommand*\natexlab[1]{#1}
\providecommand*\mciteSetBstSublistMode[1]{}
\providecommand*\mciteSetBstMaxWidthForm[2]{}
\providecommand*\mciteBstWouldAddEndPuncttrue
  {\def\EndOfBibitem{\unskip.}}
\providecommand*\mciteBstWouldAddEndPunctfalse
  {\let\EndOfBibitem\relax}
\providecommand*\mciteSetBstMidEndSepPunct[3]{}
\providecommand*\mciteSetBstSublistLabelBeginEnd[3]{}
\providecommand*\EndOfBibitem{}
\mciteSetBstSublistMode{f}
\mciteSetBstMaxWidthForm{subitem}{(\alph{mcitesubitemcount})}
\mciteSetBstSublistLabelBeginEnd
  {\mcitemaxwidthsubitemform\space}
  {\relax}
  {\relax}

\bibitem[Tadmor and Miller(2012)Tadmor, and Miller]{Tadmor_1}
Tadmor,~E.~B.; Miller,~R.~E. \emph{Modeling Materials: Continuum, Atomistic and
  Multiscale Techniques}; Cambridge University Press, 2012; pp 153--300\relax
\mciteBstWouldAddEndPuncttrue
\mciteSetBstMidEndSepPunct{\mcitedefaultmidpunct}
{\mcitedefaultendpunct}{\mcitedefaultseppunct}\relax
\EndOfBibitem
\bibitem[Hautier \latin{et~al.}(2012)Hautier, Jain, and Ong]{Hautier_1}
Hautier,~G.; Jain,~A.; Ong,~S.~P. \emph{J. Mater. Sci.} \textbf{2012},
  \emph{47}, 7317--7340\relax
\mciteBstWouldAddEndPuncttrue
\mciteSetBstMidEndSepPunct{\mcitedefaultmidpunct}
{\mcitedefaultendpunct}{\mcitedefaultseppunct}\relax
\EndOfBibitem
\bibitem[Burke(2012)]{Burke_1}
Burke,~K. \emph{J. Chem. Phys.} \textbf{2012}, \emph{136}, 150901\relax
\mciteBstWouldAddEndPuncttrue
\mciteSetBstMidEndSepPunct{\mcitedefaultmidpunct}
{\mcitedefaultendpunct}{\mcitedefaultseppunct}\relax
\EndOfBibitem
\bibitem[Neugebauer and Hickel(2013)Neugebauer, and Hickel]{Neugebauer_1}
Neugebauer,~J.; Hickel,~T. \emph{Comp. Mol. Sci.} \textbf{2013}, \emph{3},
  438--448\relax
\mciteBstWouldAddEndPuncttrue
\mciteSetBstMidEndSepPunct{\mcitedefaultmidpunct}
{\mcitedefaultendpunct}{\mcitedefaultseppunct}\relax
\EndOfBibitem
\bibitem[Hill \latin{et~al.}(2007)Hill, Freeman, and Subramanian]{Hill_1}
Hill,~J.-R.; Freeman,~C.~M.; Subramanian,~L. \emph{Use of Force Fields in
  Materials Modeling}; John Wiley and Sons, Inc., 2007; pp 141--216\relax
\mciteBstWouldAddEndPuncttrue
\mciteSetBstMidEndSepPunct{\mcitedefaultmidpunct}
{\mcitedefaultendpunct}{\mcitedefaultseppunct}\relax
\EndOfBibitem
\bibitem[Torrens(1972)]{Torrens_1}
Torrens,~I.~M. \emph{Interatomic potentials}; Academic Press Inc., 1972\relax
\mciteBstWouldAddEndPuncttrue
\mciteSetBstMidEndSepPunct{\mcitedefaultmidpunct}
{\mcitedefaultendpunct}{\mcitedefaultseppunct}\relax
\EndOfBibitem
\bibitem[Elliott(2011)]{Elliott_1}
Elliott,~J.~A. \emph{Int. Mat. Rev.} \textbf{2011}, \emph{56}, 207\relax
\mciteBstWouldAddEndPuncttrue
\mciteSetBstMidEndSepPunct{\mcitedefaultmidpunct}
{\mcitedefaultendpunct}{\mcitedefaultseppunct}\relax
\EndOfBibitem
\bibitem[Bianchini \latin{et~al.}(2016)Bianchini, Kermode, and
  Vita]{Bianchini_1}
Bianchini,~F.; Kermode,~J.~R.; Vita,~A.~D. \emph{Modelling Simul. Mater. Sci.
  Eng.} \textbf{2016}, \emph{24}, 045012\relax
\mciteBstWouldAddEndPuncttrue
\mciteSetBstMidEndSepPunct{\mcitedefaultmidpunct}
{\mcitedefaultendpunct}{\mcitedefaultseppunct}\relax
\EndOfBibitem
\bibitem[Witten \latin{et~al.}(2011)Witten, Frank, and Hall]{Witten_1}
Witten,~I.~H.; Frank,~E.; Hall,~M.~A. \emph{Data mining: Practical machine
  learning tools and techniques}; Elsevier, 2011\relax
\mciteBstWouldAddEndPuncttrue
\mciteSetBstMidEndSepPunct{\mcitedefaultmidpunct}
{\mcitedefaultendpunct}{\mcitedefaultseppunct}\relax
\EndOfBibitem
\bibitem[Hastie \latin{et~al.}(2009)Hastie, Tibshirani, and Friedman]{Hastie_1}
Hastie,~T.; Tibshirani,~R.; Friedman,~J. \emph{The Elements of Statistical
  Learning: Data Mining, Inference, and Prediction}, 2nd ed.; Springer: New
  York, 2009\relax
\mciteBstWouldAddEndPuncttrue
\mciteSetBstMidEndSepPunct{\mcitedefaultmidpunct}
{\mcitedefaultendpunct}{\mcitedefaultseppunct}\relax
\EndOfBibitem
\bibitem[Hofmann \latin{et~al.}(2008)Hofmann, Scholkopf, and Smola]{Hoffman_1}
Hofmann,~T.; Scholkopf,~B.; Smola,~A.~J. \emph{Ann. Statist.} \textbf{2008},
  \emph{36}, 1171\relax
\mciteBstWouldAddEndPuncttrue
\mciteSetBstMidEndSepPunct{\mcitedefaultmidpunct}
{\mcitedefaultendpunct}{\mcitedefaultseppunct}\relax
\EndOfBibitem
\bibitem[Behler(2011)]{Behler_1}
Behler,~J. \emph{J. Chem. Phys.} \textbf{2011}, \emph{134}, 074106\relax
\mciteBstWouldAddEndPuncttrue
\mciteSetBstMidEndSepPunct{\mcitedefaultmidpunct}
{\mcitedefaultendpunct}{\mcitedefaultseppunct}\relax
\EndOfBibitem
\bibitem[Bart\'ok \latin{et~al.}(2010)Bart\'ok, Payne, Kondor, and
  Cs\'anyi]{Bartok_2}
Bart\'ok,~A.~P.; Payne,~M.~C.; Kondor,~R.; Cs\'anyi,~G. \emph{Phys. Rev. Lett.}
  \textbf{2010}, \emph{104}, 136403\relax
\mciteBstWouldAddEndPuncttrue
\mciteSetBstMidEndSepPunct{\mcitedefaultmidpunct}
{\mcitedefaultendpunct}{\mcitedefaultseppunct}\relax
\EndOfBibitem
\bibitem[Lorenz \latin{et~al.}(2004)Lorenz, Gro{\ss}, and Scheffler]{Lorenz_1}
Lorenz,~S.; Gro{\ss},~A.; Scheffler,~M. \emph{Chem. Phys. Lett.} \textbf{2004},
  \emph{395}, 210 -- 215\relax
\mciteBstWouldAddEndPuncttrue
\mciteSetBstMidEndSepPunct{\mcitedefaultmidpunct}
{\mcitedefaultendpunct}{\mcitedefaultseppunct}\relax
\EndOfBibitem
\bibitem[Botu and Ramprasad(2015)Botu, and Ramprasad]{Botu_1}
Botu,~V.; Ramprasad,~R. \emph{Int. J. Quant. Chem.} \textbf{2015}, \emph{115},
  1074--1083\relax
\mciteBstWouldAddEndPuncttrue
\mciteSetBstMidEndSepPunct{\mcitedefaultmidpunct}
{\mcitedefaultendpunct}{\mcitedefaultseppunct}\relax
\EndOfBibitem
\bibitem[Botu and Ramprasad(2015)Botu, and Ramprasad]{Botu_2}
Botu,~V.; Ramprasad,~R. \emph{Phys. Rev. B} \textbf{2015}, \emph{92},
  094306\relax
\mciteBstWouldAddEndPuncttrue
\mciteSetBstMidEndSepPunct{\mcitedefaultmidpunct}
{\mcitedefaultendpunct}{\mcitedefaultseppunct}\relax
\EndOfBibitem
\bibitem[Li \latin{et~al.}(2015)Li, Kermode, and De~Vita]{Li_1}
Li,~Z.; Kermode,~J.~R.; De~Vita,~A. \emph{Phys. Rev. Lett.} \textbf{2015},
  \emph{114}, 096405\relax
\mciteBstWouldAddEndPuncttrue
\mciteSetBstMidEndSepPunct{\mcitedefaultmidpunct}
{\mcitedefaultendpunct}{\mcitedefaultseppunct}\relax
\EndOfBibitem
\bibitem[Kohn and Sham(1965)Kohn, and Sham]{Kohn_1}
Kohn,~W.; Sham,~L.~J. \emph{Phys. Rev.} \textbf{1965}, \emph{140},
  A1133--A1138\relax
\mciteBstWouldAddEndPuncttrue
\mciteSetBstMidEndSepPunct{\mcitedefaultmidpunct}
{\mcitedefaultendpunct}{\mcitedefaultseppunct}\relax
\EndOfBibitem
\bibitem[Hohenberg and Kohn(1964)Hohenberg, and Kohn]{Hohenberg_1}
Hohenberg,~P.; Kohn,~W. \emph{Phys. Rev.} \textbf{1964}, \emph{136},
  B864--B871\relax
\mciteBstWouldAddEndPuncttrue
\mciteSetBstMidEndSepPunct{\mcitedefaultmidpunct}
{\mcitedefaultendpunct}{\mcitedefaultseppunct}\relax
\EndOfBibitem
\bibitem[Car and Parrinello(1985)Car, and Parrinello]{Car_1}
Car,~R.; Parrinello,~M. \emph{Phys. Rev. Lett.} \textbf{1985}, \emph{55},
  2471--2474\relax
\mciteBstWouldAddEndPuncttrue
\mciteSetBstMidEndSepPunct{\mcitedefaultmidpunct}
{\mcitedefaultendpunct}{\mcitedefaultseppunct}\relax
\EndOfBibitem
\bibitem[Kresse and Furthmuller(1996)Kresse, and Furthmuller]{Kresse1}
Kresse,~G.; Furthmuller,~J. \emph{Phys. Rev. B} \textbf{1996}, \emph{54},
  11169\relax
\mciteBstWouldAddEndPuncttrue
\mciteSetBstMidEndSepPunct{\mcitedefaultmidpunct}
{\mcitedefaultendpunct}{\mcitedefaultseppunct}\relax
\EndOfBibitem
\bibitem[Kresse and Joubert(1999)Kresse, and Joubert]{Kresse2}
Kresse,~G.; Joubert,~D. \emph{Phys. Rev. B} \textbf{1999}, \emph{59},
  1758\relax
\mciteBstWouldAddEndPuncttrue
\mciteSetBstMidEndSepPunct{\mcitedefaultmidpunct}
{\mcitedefaultendpunct}{\mcitedefaultseppunct}\relax
\EndOfBibitem
\bibitem[Perdew \latin{et~al.}(1996)Perdew, Burke, and Wang]{Perdew1}
Perdew,~J.~P.; Burke,~K.; Wang,~Y. \emph{Phys. Rev. B} \textbf{1996},
  \emph{54}, 16533\relax
\mciteBstWouldAddEndPuncttrue
\mciteSetBstMidEndSepPunct{\mcitedefaultmidpunct}
{\mcitedefaultendpunct}{\mcitedefaultseppunct}\relax
\EndOfBibitem
\bibitem[Bl\"{o}chl(1994)]{Blochl1}
Bl\"{o}chl,~P.~E. \emph{Phys. Rev. B} \textbf{1994}, \emph{50}, 17953\relax
\mciteBstWouldAddEndPuncttrue
\mciteSetBstMidEndSepPunct{\mcitedefaultmidpunct}
{\mcitedefaultendpunct}{\mcitedefaultseppunct}\relax
\EndOfBibitem
\bibitem[Bart\'ok \latin{et~al.}(2013)Bart\'ok, Kondor, and Cs\'anyi]{Bartok_1}
Bart\'ok,~A.~P.; Kondor,~R.; Cs\'anyi,~G. \emph{Phys. Rev. B} \textbf{2013},
  \emph{87}, 184115\relax
\mciteBstWouldAddEndPuncttrue
\mciteSetBstMidEndSepPunct{\mcitedefaultmidpunct}
{\mcitedefaultendpunct}{\mcitedefaultseppunct}\relax
\EndOfBibitem
\bibitem[Jolliffe(2014)]{Jolliffe_1}
Jolliffe,~I. \emph{Statistics Reference Online}; John Wiley and Sons, Inc.,
  2014\relax
\mciteBstWouldAddEndPuncttrue
\mciteSetBstMidEndSepPunct{\mcitedefaultmidpunct}
{\mcitedefaultendpunct}{\mcitedefaultseppunct}\relax
\EndOfBibitem
\bibitem[Sch{\"o}lkopf \latin{et~al.}(1997)Sch{\"o}lkopf, Smola, and
  M{\"u}ller]{Scholkopf_1}
Sch{\"o}lkopf,~B.; Smola,~A.; M{\"u}ller,~K.-R. Kernel principal component
  analysis. International Conference on Artificial Neural Networks. 1997; pp
  583--588\relax
\mciteBstWouldAddEndPuncttrue
\mciteSetBstMidEndSepPunct{\mcitedefaultmidpunct}
{\mcitedefaultendpunct}{\mcitedefaultseppunct}\relax
\EndOfBibitem
\bibitem[Cox and Cox(2001)Cox, and Cox]{Cox_1}
Cox,~T.~F.; Cox,~M.~A. Modern multidimensional scaling. 2001\relax
\mciteBstWouldAddEndPuncttrue
\mciteSetBstMidEndSepPunct{\mcitedefaultmidpunct}
{\mcitedefaultendpunct}{\mcitedefaultseppunct}\relax
\EndOfBibitem
\bibitem[Behler(2011)]{Behler_2}
Behler,~J. \emph{Phys. Chem. Chem. Phys.} \textbf{2011}, \emph{13}, 17930\relax
\mciteBstWouldAddEndPuncttrue
\mciteSetBstMidEndSepPunct{\mcitedefaultmidpunct}
{\mcitedefaultendpunct}{\mcitedefaultseppunct}\relax
\EndOfBibitem
\bibitem[Muller \latin{et~al.}(2001)Muller, Mika, Ratsch, Tsuda, and
  Scholkopf]{Muller_1}
Muller,~K.~R.; Mika,~S.; Ratsch,~G.; Tsuda,~K.; Scholkopf,~B. \emph{IEEE Trans.
  Neural Networks} \textbf{2001}, \emph{12}, 181\relax
\mciteBstWouldAddEndPuncttrue
\mciteSetBstMidEndSepPunct{\mcitedefaultmidpunct}
{\mcitedefaultendpunct}{\mcitedefaultseppunct}\relax
\EndOfBibitem
\bibitem[Hansen \latin{et~al.}(2013)Hansen, Montavon, Biegler, Fazil, Rupp,
  Scheffler, von Lilienfeld, Tkatchenko, and Muller]{Hansen_1}
Hansen,~K.; Montavon,~G.; Biegler,~F.; Fazil,~S.; Rupp,~M.; Scheffler,~M.; von
  Lilienfeld,~O.~A.; Tkatchenko,~A.; Muller,~K. \emph{J. Chem. Theory Comput.}
  \textbf{2013}, \emph{9}, 3404\relax
\mciteBstWouldAddEndPuncttrue
\mciteSetBstMidEndSepPunct{\mcitedefaultmidpunct}
{\mcitedefaultendpunct}{\mcitedefaultseppunct}\relax
\EndOfBibitem
\bibitem[Rupp(2015)]{Rupp_1}
Rupp,~M. \emph{Int. J. Quant. Chem.} \textbf{2015}, \emph{115},
  1058--1073\relax
\mciteBstWouldAddEndPuncttrue
\mciteSetBstMidEndSepPunct{\mcitedefaultmidpunct}
{\mcitedefaultendpunct}{\mcitedefaultseppunct}\relax
\EndOfBibitem
\bibitem[Jacobsen \latin{et~al.}(1987)Jacobsen, Norskov, and Puska]{Jacobsen_1}
Jacobsen,~K.~W.; Norskov,~J.~K.; Puska,~M.~J. \emph{Phys. Rev. B}
  \textbf{1987}, \emph{35}, 7423--7442\relax
\mciteBstWouldAddEndPuncttrue
\mciteSetBstMidEndSepPunct{\mcitedefaultmidpunct}
{\mcitedefaultendpunct}{\mcitedefaultseppunct}\relax
\EndOfBibitem
\bibitem[Plimpton(1995)]{Plimpton_1}
Plimpton,~S. \emph{J. Comp. Phys.} \textbf{1995}, \emph{117}, 1--19\relax
\mciteBstWouldAddEndPuncttrue
\mciteSetBstMidEndSepPunct{\mcitedefaultmidpunct}
{\mcitedefaultendpunct}{\mcitedefaultseppunct}\relax
\EndOfBibitem
\bibitem[Alavi and Thompson(2006)Alavi, and Thompson]{Saman_1}
Alavi,~S.; Thompson,~D.~L. \emph{J. Phys. Chem. A} \textbf{2006}, \emph{110},
  1518--1523, PMID: 16435812\relax
\mciteBstWouldAddEndPuncttrue
\mciteSetBstMidEndSepPunct{\mcitedefaultmidpunct}
{\mcitedefaultendpunct}{\mcitedefaultseppunct}\relax
\EndOfBibitem
\bibitem[Neyts and Bogaerts(2009)Neyts, and Bogaerts]{Erik_1}
Neyts,~E.~C.; Bogaerts,~A. \emph{J. Phys. Chem. C} \textbf{2009}, \emph{113},
  2771--2776\relax
\mciteBstWouldAddEndPuncttrue
\mciteSetBstMidEndSepPunct{\mcitedefaultmidpunct}
{\mcitedefaultendpunct}{\mcitedefaultseppunct}\relax
\EndOfBibitem
\bibitem[Zhang \latin{et~al.}(2007)Zhang, Stocks, and Zhong]{Zhang_1}
Zhang,~K.; Stocks,~G.~M.; Zhong,~J. \emph{Nanotechnology} \textbf{2007},
  \emph{18}, 285703\relax
\mciteBstWouldAddEndPuncttrue
\mciteSetBstMidEndSepPunct{\mcitedefaultmidpunct}
{\mcitedefaultendpunct}{\mcitedefaultseppunct}\relax
\EndOfBibitem
\bibitem[Zhou \latin{et~al.}(2002)Zhou, Karplus, Ball, and Berry]{Zhou_1}
Zhou,~Y.; Karplus,~M.; Ball,~K.~D.; Berry,~R.~S. \emph{J. Chem. Phys.}
  \textbf{2002}, \emph{116}, 2323--2329\relax
\mciteBstWouldAddEndPuncttrue
\mciteSetBstMidEndSepPunct{\mcitedefaultmidpunct}
{\mcitedefaultendpunct}{\mcitedefaultseppunct}\relax
\EndOfBibitem
\bibitem[Stoltze \latin{et~al.}(1988)Stoltze, N\o{}rskov, and
  Landman]{Stoltze_1}
Stoltze,~P.; N\o{}rskov,~J.~K.; Landman,~U. \emph{Phys. Rev. Lett.}
  \textbf{1988}, \emph{61}, 440--443\relax
\mciteBstWouldAddEndPuncttrue
\mciteSetBstMidEndSepPunct{\mcitedefaultmidpunct}
{\mcitedefaultendpunct}{\mcitedefaultseppunct}\relax
\EndOfBibitem
\bibitem[Pham \latin{et~al.}(2011)Pham, Williams, Mahaffey, Radovic, Arroyave,
  and Cagin]{Pham_1}
Pham,~H.~H.; Williams,~M.~E.; Mahaffey,~P.; Radovic,~M.; Arroyave,~R.;
  Cagin,~T. \emph{Phys. Rev. B} \textbf{2011}, \emph{84}, 064101\relax
\mciteBstWouldAddEndPuncttrue
\mciteSetBstMidEndSepPunct{\mcitedefaultmidpunct}
{\mcitedefaultendpunct}{\mcitedefaultseppunct}\relax
\EndOfBibitem
\end{mcitethebibliography}

\end{document}